%% file: Manual-arXiv-v2.tex
\documentclass[a4paper,10pt,twoside]{report}

\input{PackagesnStyles.tex}

\graphicspath{{Figures/}}

\input{Definitions.tex}

\allowdisplaybreaks

\begin{document}

% (Create a counter)
\newcounter{tblEqCounter}     

% ====================================== Page Numbering
\pagestyle{plain}
\hypersetup{pageanchor=false}
% ====================================== Page Numbering II
\hypersetup{pageanchor=true}
\pagenumbering{roman}
\setcounter{page}{1}
% ====================================== Page Numbering III
\pagenumbering{arabic}
\setcounter{page}{1}

\begin{center}
{\Large FeynMaster: a plethora of Feynman tools}

\vspace{4mm}

\renewcommand*{\thefootnote}{\fnsymbol{footnote}}

Duarte Fontes\fn{duartefontes@tecnico.ulisboa.pt}, Jorge C. Romão\fn{jorge.romao@tecnico.ulisboa.pt}

\renewcommand*{\thefootnote}{\arabic{footnote}}
\setcounter{footnote}{0}

\vspace{1mm}

\textit{Departamento de Física and CFTP, Instituto Superior Técnico\\[-2mm]
Universidade de Lisboa, Av. Rovisco Pais 1, 1049-001 Lisboa, Portugal,} 

(Dated: \today)

\vspace{1mm}
\end{center}

\vs{1.2mm}
\begin{addmargin}[12mm]{12mm}
	\small
	\textsc{FeynMaster} is a multi-tasking software for particle physics studies. By making use of already existing programs (\textsc{FeynRules}, \textsc{QGRAF}, \textsc{FeynCalc}), \textsc{FeynMaster} automatically  generates Feynman rules, generates and draws Feynman diagrams, generates amplitudes, performs both loop and algebraic calculations, and fully renormalizes models. In parallel with this automatic character, \textsc{FeynMaster} allows the user to manipulate the generated results in \texttt{\textsc{Mathematica}} notebooks in a flexible and consistent way. It can be downloaded in \url{https://porthos.tecnico.ulisboa.pt/FeynMaster/}.
\end{addmargin}

\normalsize

\section{Introduction}
\label{sec:Intro}

\n The Large Hadron Collider (LHC) has completed its second run, and no physics beyond the Standard Model (BSM) has been found. In face of such absence of experimental guidance, a great diversity of BSM models are intensively studied today. As they must explain recent data from the LHC, they are required to make precise predictions. This inevitably forces model builders to consider next-to-leading-order (NLO) corrections, which in turn demand a sound renormalization program to treat {ultraviolet (UV)} divergences. Given the complexity of NLO and renormalization calculations, the use of computational tools is today virtually indispensable. In the specific context of model building and NLO calculations, there is currently a myriad of softwares addressing one or several of the following tasks (e.g. refs. \cite{Belanger:2003sd,Cullen:2011ac,Cullen:2014yla,Lorca:2004fg,Degrande:2014vpa,FeynCalc-new,FeynCalc-old,FeynRules-new,FeynRules-old,Hahn:1998yk,Hahn:2000kx,Kublbeck:1990xc,Pukhov:1999gg,QGRAF,Semenov:1996es,Semenov:1998eb,Tentyukov:1999is,Wang:2004du,Alwall:2014hca}):
\vs{1.0mm}
\begin{center}
	\quad a) generation of Feynman rules; \qquad b) generation and drawing of Feynman diagrams;\\
	c) generation of amplitudes; \hs{3mm} d) loop calculations; \hs{3mm} e) algebraic calculations; \hs{3mm} f) renormalization.
\end{center}
\vs{1.0mm}

\n However, despite the undisputed quality of some of the referred softwares --- which perform almost all {tasks} of the above list ---, they usually do not combine an automatic character with the possibility of manipulating the final analytical expressions in a practical way. And although interfaces between different softwares exist, they tend not to be free of constraints, since the notation changes between softwares and a conversion is not totally automatic. It would thus be desirable to have a single software that could perform all the above listed tasks, and at the same time allowing the user to handle the final results.

\n In this paper, a new such software is introduced. \ts{FeynMaster} is a single program, written in both \t{\ts{Python}} and \t{\ts{Mathematica}}, that combines \ts{FeynRules}~\cite{FeynRules-old,FeynRules-new}, \ts{QGRAF}~\cite{QGRAF} and \ts{FeynCalc}~\cite{FeynCalc-old,FeynCalc-new} to perform \textit{all} the referred tasks in a flexible and consistent way. We highlight four of its major advantages. First, \ts{FeynMaster} has a hybrid character concerning automatization: not only does it automatically generate the results, but it also allows the user to act upon them. This feature is extremely useful, since very often in research one is not interested in obtaining a rigid list of final expressions, but in handling them at will. This is made possible in \ts{FeynMaster} due to the creation of notebooks for both \ts{FeynRules} and \ts{FeynCalc}, in which a multiplicity of different tools enables the user to manipulate the results. Second, the complete set of analytical expressions for the counterterms in the modified minimal subtraction ($\overline{\text{MS}}$) scheme can be automatically calculated. Third, \ts{FeynMaster} includes a thorough interaction with numerical calculations, as it converts the expressions to \ts{LoopTools}~\cite{Hahn:1998yk} form and writes a complete \t{\ts{Fortran}} program to use them. Finally, all the printable outputs of \ts{FeynMaster} --- the complete set of Feynman rules (both for non-renormalizable interactions and counterterms interactions), the Feynman diagrams, as well as a list containing both the expressions and computed counterterms --- are written in \LaTeX \, files, which makes their inclusion in a paper remarkably simple.

\n The general usage of \ts{FeynMaster} can be summarized in a few lines: after the user has defined the model%
%
%(with both a regular \ts{FeynRules} model file and a \ts{FeynMaster}-specific model file)
%
, the \ts{FeynMaster} run is controlled from a single file (\t{Control.py}). Here, the sequence of processes to study can be chosen, as well as many different options. Once \t{Control.py} is edited, \ts{FeynMaster} is ready to run. The run automatically generates and opens several PDF files --- according to the options chosen in \t{Control.py} --- and creates the above-mentioned notebooks.

\n This paper is organized as follows. In section \ref{sec:Instal}, we explain how to download and install \ts{FeynMaster}. Section \ref{sec:Create} is devoted to the creation of models, and section \ref{sec:Usage} to the detailed usage of \ts{FeynMaster}. Then, in section \ref{sec:Examples}, we give some examples{, and in section 
	\ref{sec:Comparison} we briefly compare \ts{FeynMaster} with other softwares. Finally, we present a brief summary of the paper in section \ref{sec:Summary}, as well as a quick first usage of \ts{FeynMaster}.}

%was not built to compete with other software, but to unify different tools performed by already existing software.

% 

% \ts{FeynMaster} is a software that combines model-building with loop calculations, yielding both analytical and numerical results, with the flexibility to allow control over the results. 
%
%Practical purposes, some cases analytical, some cases numerical results. In the former, it is desirable to have not only the final results in a automatic fashion, but also to have control over those results.

\n

\section{Installation}
\label{sec:Instal}

\subsection{Download}
\label{sec:Download}

\n \ts{FeynMaster} can be downloaded in:
\begin{center}
	\url{https://porthos.tecnico.ulisboa.pt/FeynMaster/}.
\end{center}
\ts{FeynRules}, \ts{QGRAF} and \ts{FeynCalc}, essential to run \ts{FeynMaster}, can be downloaded in \url{https://feynrules.irmp.ucl.ac.be/}, \url{http://cfif.ist.utl.pt/~paulo/qgraf.html} and \url{https://feyncalc.github.io/}, respectively.%
\fn{The user of \ts{FeynMaster} is required to be familiar with both \ts{FeynRules} (in order to define new models) and \ts{FeynCalc} (in order to manipulate the final results). In a first approach, there is really no need to learn QGRAF, since the most non-trivial part of this program --- the definition of the model --- is automatically carried through by \ts{FeynMaster}.
	To run \ts{FeynMaster}, it is also necessary to have \t{\ts{Python}}, \t{\ts{Mathematica}} and \LaTeX \, installed; links to download are \url{https://www.python.org/downloads/} \url{http://www.wolfram.com/mathematica/} and \url{https://www.latex-project.org/get/}, respectively.
	We tested \ts{FeynMaster} using version 3.6 of \t{\ts{Python}} and version {12.0} of \t{\ts{Mathematica}}. %
	(Note that \ts{FeynMaster} will \textit{not} run if a \t{\ts{Python}} version prior to 3 is used{, and it is only guaranteed to properly work if a version of \t{\ts{Mathematica}} not older than 10.3 is used}.)
	As for \LaTeX \,, the user is required to update the package database; note also that, in the first run of \ts{FeynMaster}, some packages (like \t{feynmp-auto} and \t{breqn}) may require authorization to be installed.
	We verified that \ts{FeynMaster} runs properly with the latest public versions of \ts{FeynRules}, \ts{QGRAF} and \ts{FeynCalc} --- namely, versions 2.3.34, 3.4.2 and 9.3.0, respectively. More instructions on how to download and install QGRAF can be found in \url{https://porthos.tecnico.ulisboa.pt/CTQFT/node9.html} and in \url{https://porthos.tecnico.ulisboa.pt/CTQFT/node33.html}. Finally, note that, when using Linux or Mac, the executable QGRAF file should be named {\t{qgraf}}.}
After downloading \ts{FeynMaster}, the downloaded file should be extracted, and the resulting folder (named `\ts{FeynMaster}') can be placed in any directory of the user's choice.

\subsection{Defining the directories}
\label{sec:direct}

\n Once the `\ts{FeynMaster}' folder is set in a specific directory, further directories should be specified. This must be done by editing two files inside that folder: the \t{RUN-FeynMaster} batch\fn{This file will have different formats according to the operative system: \t{.bat} in Windows and \t{.sh} in Linux/Mac.} and \t{Control.py}. For the former, the user should edit the file simply by completing the \t{(...)} in the second line with the directory where the \ts{FeynMaster} folder is. For the latter, \t{Control.py} must be opened; then, in the last section (\textit{Directories section}), and for each one of the variables defined there (\t{dirFM}, \t{dirFR}, ..., \t{dirCT}), the user should specify a path.\fn{Note that, while in Linux/Mac a (single) forward slash should be used to separate the different levels of the directory, Windows requires a double backslash.} The meaning of the different variables is explained in Table \ref{tab:dirs}.\fn{We should clarify the difference between \textit{the directory corresponding to x} and \textit{the directory where x is}: while in the former \textit{x} itself is included in the directory path, in the latter it is not. For example, suppose that the folder \textit{x} lies inside the directory `documents'; then, \textit{the directory corresponding to x} means `/documents/x/', while \textit{the directory where x is} means `/documents/'.}
\begin{table}[!h]%
\begin{normalsize}
\normalsize
\begin{center}
\begin{tabular}
%{@{}p{3cm} l @{}}
{@{\hspace{3mm}}>{\raggedright\arraybackslash}p{2cm}>{\raggedright\arraybackslash}p{11.8cm}@{\hspace{3mm}}}
\hlinewd{1.1pt}
Variable & Meaning \\
\hline\\[-1.5mm]
\t{dirFM} & directory corresponding to \ts{FeynMaster} \\[2.5mm]
\t{dirFR} & directory corresponding to \ts{FeynRules} \\[2.5mm]
\t{dirFRmod} &directory where the \ts{FeynRules} model is \\[2.5mm]
\t{dirQ} &  directory corresponding to \ts{QGRAF} \\[2.5mm]
\t{dirQmod} & directory where the \ts{QGRAF} model is to be found \\[2.5mm]
\t{dirMain} & directory corresponding to \ts{FeynMaster} output \\[2.5mm]
\t{dirPro} & directory corresponding to the one where processes are to be stored \\[2.5mm]
\t{dirFey} & directory corresponding to the one where the non-renormalized Feynman rules are to be stored \\[0.5mm]
\t{dirCT} & directory corresponding to the one where the counterterms are to be stored \\[2.5mm]
\hlinewd{1.1pt}
\end{tabular}
\end{center}
\vspace{-5mm}
\end{normalsize}
\caption{Directories to be specified in the end of \t{Control.py}. See text for details.}
\label{tab:dirs}
\end{table}
\normalsize

\n Some additional remarks are in order. First, the tilde ($\textasciitilde$) symbol should not be used to simplify directories. Second, the folder `FeynRulesModels' inside the `\ts{FeynMaster}' folder should be cut and pasted inside the directory corresponding to \ts{FeynRules}.
Third, the user must create an empty folder named `ModelsFeynMaster' inside \t{dirQ} (such folder corresponds to \t{dirQmod}, and is the place where the generated QGRAF model files will be placed).
Fourth, the directories may depend on the specific model being used at the time; in those cases, the user can use \t{\ts{Python}} features to write the string as depending on other variables.\fn{By default, \t{Control.py} already uses all these features, so that the user only needs to fill in the \t{(...)} parts. In any event, let us illustrate what is being said with an example. The directory where the \ts{FeynRules} model is should be specified as a function of the variable \t{intname} (defined in the \ts{FeynMaster} model, see below); for instance, in Linux,
	\vspace{-1mm}
	\begin{center}
		\t{dirFRmod = `home/programs/\ts{FeynRules}/Models/' + intname + `/'}.
	\end{center}
	\vspace{-1mm}
	Note also that this feature --- writing strings as depending on other variables --- can be used to simplify a directory which lies inside a previously defined directory. For example, if the directory where the \ts{FeynRules} model is lies inside the directory corresponding to \ts{FeynRules}, then we can write, for example,
	\vspace{-1mm}
	\begin{center}
		\t{dirFRmod = dirFR + `Models/' + intname + `/'}.
\end{center}}
Finally, the last four directories --- \t{dirMain}, \t{dirPro}, \t{dirFey}, \t{dirCT} --- do not need to exist before the user defines them in the \t{Control.py} file, for they are automatically created.

\n To test \ts{FeynMaster}, jump to section \ref{sec:Summary}, where instructions for a quick first usage are given.

\section{Creating a new model}
\label{sec:Create}

\n \ts{FeynMaster} is a model dependent program: it cannot run without the specification of a model. For \textit{each model}, such specification corresponds to the definition of \textit{two files}: the \ts{FeynRules} model file and the \ts{FeynMaster} model file. In this section, we explain in detail how to create both; a concrete example is given in section \ref{sec:FullReno}. Note that \ts{FeynMaster} already comes with three models (and hence six files): QED, Scalar QED (SQED) and the Standard Model.\fn{The Standard Model \ts{FeynRules} model file is written in an arbitrary $R_{\xi}$ gauge and with the $\eta$ parameters of ref.~\cite{Romao:2012pq}, while the \ts{FeynMaster} model file closely follows ref.~\cite{Denner:1991kt} to define the renormalization.}
These serve as prototypes, and we highly recommend them as guiding tools in the creation of a new model.

\n Since there are already models available, this section can be skipped in a first utilization of \ts{FeynMaster}.

\subsection{Defining the \ts{FeynRules} model}

\n The \ts{FeynRules} model is an usual model for \ts{FeynRules}.\fn{It is assumed here that the user is familiar with \ts{FeynRules}, and knows how to create a \ts{FeynRules} model. If this is not the case, we refer to the \ts{FeynRules} website \url{https://feynrules.irmp.ucl.ac.be/}.}  In order for \ts{FeynMaster} to be able to work with it, the final Lagrangean should be included in the \ts{FeynRules} model. Moreover, this Lagrangean must be separated in different parts --- each of them corresponding to a different type of interaction ---, in such a way that each of those parts should have a specific name: see Table \ref{tab:FRnames}.\fn{To be clear, it is not mandatory to define all the 6 Lagrangean parts present in Table \ref{tab:FRnames} (for example, if the model does not have a ghost sector, there is no need to define LGhost); only, no other name besides the 6 ones specified in the right column of Table \ref{tab:FRnames} will be recognized by \ts{FeynMaster}.}
\begin{table}[!h]%
\begin{normalsize}
\normalsize
\begin{center}
\begin{tabular}
%{@{}p{3cm} l @{}}
{@{\hspace{3mm}}>{\raggedright\arraybackslash}p{5cm}>{\raggedright\arraybackslash}p{2.9cm}@{\hspace{3mm}}}
\hlinewd{1.1pt}
Type of interactions & \ts{FeynRules} name \\
\hline
pure gauge & \t{LGauge} \\
fermion-gauge & \t{LFermions}\\
Yukawa & \t{LYukawa} \\
scalar-scalar and gauge-scalar & \t{LHiggs} \\
ghosts & \t{LGhost} \\
gauge fixing & \t{LGF} \\
\hlinewd{1.1pt}
\end{tabular}
\end{center}
%\end{footnotesize}
\vspace{-5mm}
%\end{large}
\end{normalsize}
\caption{Name of the different Lagrangean parts according to the type of interaction.}
\label{tab:FRnames}
\end{table}
\normalsize

\n Some additional notes on the \ts{FeynRules} model are in order. First, the Lagrangean parts involving fermions (LFermions and LYukawa) can be written either in terms of Dirac fermions or Weyl fermions; in both cases, the Feynman rules will be presented for Dirac fermions.\fn{This means that, if the user has written Lagrangean parts in terms of Weyl fermions, a relation should be specified in the particle classes between the Weyl fermions and their associated Dirac spinors (see, for example, the Standard Model \ts{FeynRules} model file; for more information, cf. the \ts{FeynRules} manual).} Second, every propagating particle should have a mass assignment, either to a variable or to 0 (e.g., \t{Mass -> MH} or \t{Mass -> 0}). Finally, every parameter present in the model (non-zero masses included) should be defined in the parameters sections (\t{M\$Parameters}) with a \LaTeX \, name, so that it can be properly printed in the outputs (e.g., \t{TeX -> Subscript[m, h]}). A value can also be associated to each parameter (e.g., \t{Value -> 125});\fn{In the case of matricial parameters, the variable name is \t{Values} instead of \t{Value}, and the set of values should be written as a \t{\ts{Mathematica}} list. For example, for a parameter with dimensions $2\times2$, we can set \t{Values -> \{0.5, 1, 1.5, 2\}}, for the matrix entries 11, 12, 21, 22, respectively. The Standard Model file includes values according to \cite{Tanabashi:2018oca}.} this is useful only in case the user wants to exploit the numerical interface of \ts{FeynMaster} --- to be explained in section \ref{sec:FC}.

\subsection{Defining the \ts{FeynMaster} model}

\n As we saw, besides the \ts{FeynRules} model file, \ts{FeynMaster} requires a second file: the \ts{FeynMaster} model. This is a \t{\ts{Python}} file, where further informations about the model should be specified. The different elements of the file are explained in Table \ref{tab:FMmodel}; yet, we strongly suggest the user to check some examples of \ts{FeynMaster} model files, which might be the simplest way to get acquainted with a model.
\begin{longtable}{@{\hspace{3mm}}>{\raggedright\arraybackslash}p{2.2cm}>{}p{12.8cm}@{\hspace{3mm}}}
\hlinewd{1.1pt}
Variable & Meaning \\
\hline\\[-5mm]
\t{intname} & name that shall be given to a) the \ts{QGRAF} model, which will be automatically created from the \ts{FeynRules} model, b) the folder where the \ts{FeynRules} model is to be found and c) the folder with the \ts{FeynMaster} outputs for the model at stake \\[5mm]
%\\[15mm]
\t{extname} & name that shall be printed in the \ts{FeynMaster} final documents as the model name\\[2.5mm]
\multirow{1}{15.5cm}{\centering\small{ \color{mygray}  (in the elements that follow, the user should specify the particles of the model, in \ts{QGRAF} style)}\fn{There is really no need to learn QGRAF for this: just check some examples of \ts{FeynMaster} model files. Note also that the QGRAF name of an antifermion should always be the capitalized version of the QGRAF name of the its respective fermion; the same applies for ghosts and antighosts. Some examples: for a fermion \t{f}, the antifermion should be \t{F}; for a fermion \t{e1}, the antifermion should be \t{E1}; for a ghost \t{cZ}, the antighost should be \t{CZ}.}}\\[3mm]
\t{nscl} & neutral scalars, in \ts{QGRAF} style \\[0.5mm]
\t{cscl} & charged scalars, in \ts{QGRAF} style\\[0.5mm]
\t{ancscl} & anticharged scalars, in \ts{QGRAF} style (in respective order)\\[0.5mm]
\t{fel} & fermions, in \ts{QGRAF} style\\[0.5mm]
\t{anfel} & antifermions, in \ts{QGRAF} style\\[0.5mm]
\t{ngal} & neutral gauge bosons, in \ts{QGRAF} style\\[0.5mm]
\t{cgal} & charged gauge bosons, in \ts{QGRAF} style\\[0.5mm]
\t{ancgal} & anticharged gauge bosons, in \ts{QGRAF} style (in respective order)\\[0.5mm]
\t{ngol} & neutral Goldstone bosons, in \ts{QGRAF} style\\[0.5mm]
\t{cgol} & charged Goldstone bosons, in \ts{QGRAF} style\\[0.5mm]
\t{ancgol} & anticharged Goldstone bosons, in \ts{QGRAF} style (in respective order)\\[0.5mm]
\t{ghl} & ghosts, in \ts{QGRAF} style\\[0.5mm]
\t{anghl} & antighosts, in \ts{QGRAF} style (in respective order)\\[0.5mm]
\t{glul} & gluons, in \ts{QGRAF} style\\[2.5mm]
\multirow{1}{15.5cm}{\centering\small{ \color{mygray} (in the elements that follow, the same, but in \LaTeX \,style, and in respective order)}}\\[2.5mm]
\t{LAnscl} & neutral scalars, in \LaTeX \,style \\[0.5mm]
\t{LAcscl} & charged scalars, in \LaTeX \,style\\[0.5mm]
\t{LAancscl} & anticharged scalars, in \LaTeX \,style (in respective order)\\[0.5mm]
\t{LAfel} & fermions, in \LaTeX \,style\\[0.5mm]
\t{LAanfel} & antifermions, in \LaTeX \,style\\[0.5mm]
\t{LAngal} & neutral gauge bosons, in \LaTeX \,style\\[0.5mm]
\t{LAcgal} & charged gauge bosons, in \LaTeX \,style\\[0.5mm]
\t{LAancgal} & anticharged gauge bosons, in \LaTeX \,style (in respective order)\\[0.5mm]
\t{LAngol} & neutral Goldstone bosons, in \LaTeX \,style\\[0.5mm]
\t{LAcgol} & charged Goldstone bosons, in \LaTeX \,style\\[0.5mm]
\t{LAancgol} & anticharged Goldstone bosons, in \LaTeX \,style (in respective order)\\[0.5mm]
\t{LAghl} & ghosts, in \LaTeX \,style\\[0.5mm]
\t{LAanghl} & antighosts, in \LaTeX \,style (in respective order)\\[0.5mm]
\t{LAglul} & gluons, in \LaTeX \,style\\[2.5mm]
\multirow{1}{15.5cm}{\centering\small{\color{mygray} (the elements that follow concern conventions or constraints)}}\\[2mm]
\t{FRtoQ} & set of replacement rules from the particles defined in the \ts{FeynRules} model to the particles defined in \ts{QGRAF} style\fn{There is no need to include neither the trivial replacement rules (that is, when the \ts{FeynRules} name is the same as the \ts{QGRAF} one), nor the antiparticles replacement rules.} \\[2.5mm]
\t{FRrestr} & restrictions file for the \ts{FeynRules} model\fn{The user is supposed to include all the desired simplifications for \ts{FeynRules} in the restrictions file (cf. the \ts{FeynRules} manual for more informations on restrictions files).} \\[2.5mm]
\t{FCsimp} & simplifications for \ts{FeynCalc} (in the form of replacement rules) \\[2.5mm]
\t{FCeqs} & conventions the user wants \ts{FeynCalc} to obey to (in the form of equalities) \\[2.5mm]
\t{PrMassFL} & logical variable that controls the way through which propagators are defined\fn{If \t{PrMassFL} is set to True, the poles of the propagators are extracted from the Lagrangean -- i.e., they are defined as the bilinear terms of the field at stake in the Lagrangean --, and the propagator is written in the most general form. If \t{PrMassFL} is set to False, the poles will match the variable corresponding to the mass of the propagator, and the propagator will be written in the Feynman gauge. While setting \t{PrMassFL} to True is certainly the most faithful way to describe the propagator, this option may bring certain difficulties: one the onde hand, it requires a Gauge Fixing Lagrangean to define the gauge boson propagators; on the other hand, the bilinear terms can be very complicated expressions.}\\[2.5mm]
\multirow{1}{15.5cm}{\centering\small{\color{mygray} (the remaining elements concern the renormalization of the model)}}\\[2.5mm]
\t{rensimp} & simplifications to be made immediately before renormalization\fn{This may be relevant in some cases; cf. the Standard Model \ts{FeynMaster} model file.} \\[2.0mm]
\t{renconsnum} & renormalization constants which are numbers, in \t{\ts{Mathematica}} style \\[2.0mm]
\t{LArenconsnum} & renormalization constants which are numbers, in \LaTeX \,style (in respective order) \\[2.0mm]
\t{renconsmat} & renormalization constants which are squared matrices, in \t{\ts{Mathematica}} style \\[2.0mm]
\t{LArenconsmat} & renormalization constants which are squared matrices, in \t{\ts{Mathematica}} style, in respective order \\[2.5mm]
\t{renconsmatdim} & dimension of the squared matrices just defined, in respective order \\[2.5mm]
\t{GFreno} & logical variable; when set to True, the renormalization rules are applied to the Gauge Fixing Lagrangean \\[-0.5mm]
\t{renorrules} & renormalization rules \\[1.5mm]
\hlinewd{1.1pt}
\label{tab:FMmodel}\\[-10.5mm]
\caption{Variables of the \ts{FeynMaster} model.}
\end{longtable}
\normalsize

\section{Usage}
\label{sec:Usage}

\n Once the initial specifications are concluded (i.e., once the directories, the \ts{FeynRules} model and the \ts{FeynMaster} model are defined), \ts{FeynMaster} is ready to be used. In this section, we explain in detail how the usage works. We start by showing how to edit the file that controls the \ts{FeynMaster} run. Then, after describing how to actually run \ts{FeynMaster}, we comment on its outputs, and explain how to use the two notebooks we alluded to in the Introduction.

\subsection{\t{Control.py}}
\label{sec:Control}

\n The \ts{FeynMaster} run is uniquely controlled from \t{Control.py}. In this section, we explain the different components of that file.\fn{Except the ones concerning the directories, which were addressed before. Note that a summary of the different explanations given in this section is present in the \t{Control.py} file itself as comments.} Initially, three variables should be specified, according to Table \ref{tab:inits}.
\begin{table}[!h]%
\begin{normalsize}
\normalsize
\begin{center}
\begin{tabular}
%{@{}p{3cm} l @{}}
{@{\hspace{3mm}}>{\raggedright\arraybackslash}p{1.5cm}>{\raggedright\arraybackslash}p{12.3cm}@{\hspace{3mm}}}
\hlinewd{1.1pt}
Variable & Meaning \\
\hline\\[-1.5mm]
\t{osswitch} & the operative system at stake; should be set to \t{`Windows'}, \t{`Linux'} or \t{`Mac'}\\[1.5mm]
\t{model} & the desired model; should be set to the variable \t{intname} (defined in the \ts{FeynMaster} model) of the desired model\\[4.0mm]
\t{folder} & should only be filled when the \ts{FeynMaster} model is not in the directory of \t{Control.py}, but in a sub-directory inside it; should that be the case, \t{folder} must be set to the name of such sub-directory\\
\hlinewd{1.1pt}
\end{tabular}
\end{center}
\vspace{-5mm}
\end{normalsize}
\caption{Initial variables to be defined in \t{Control.py}.}
\label{tab:inits}
\end{table}
\normalsize
While two of those variables --- \t{osswitch} and \t{folder} --- will in principle be edited once and for all in the first use of \ts{FeynMaster}, the remaining one --- \t{model} --- must be edited each time the model at stake changes.

\n After that, the user should specify the desired process (or processes). One process is specified through the definition of the set of variables shown in Table \ref{tab:process}.
\begin{table}[!h]%
\begin{normalsize}
\normalsize
\begin{center}
\begin{tabular}
%{@{}p{3cm} l @{}}
{@{\hspace{3mm}}>{\raggedright\arraybackslash}p{2.4cm}>{\raggedright\arraybackslash}p{5.7cm}>{\raggedright\arraybackslash}p{6.2cm}@{\hspace{3mm}}}
\hlinewd{1.1pt}
Variable & Meaning & Example\\
\hline\\[-1.5mm]
\t{InParticles} & incoming particles & \t{[`H']} \\[2.5mm]
\t{OutParticles} & outgoing particles & \t{[`u', `C']} \\[2.5mm]
\t{loops} & number of loops & \t{1}\\[3mm]
\t{Parsel} & selection of intermediate particles &
\multirow{3}{2.5cm}{\\[-11mm]\t{[[`avoid',`WP',`1',`3'],[`keep',} \\[-0.2mm] \t{`b',`1',`1']]}}
 \\[3.5mm]
%\t{[[`avoid',`WP',`1',`3'],[`keep',`b',`1',`1']]} \\[2.5mm]
\t{LoopTec} & loop technique in \ts{FeynCalc} & \t{3} \\[2.5mm]
\t{factor} & quantity to factor out in the end & \t{`4 Pi'} \\[2.5mm] 
\t{options} & \ts{QGRAF} options & \t{`onepi'}\\
\hlinewd{1.1pt}
\end{tabular}
\end{center}
\vspace{-5mm}
\end{normalsize}
\caption{Variables that specify one process. See text for details.}
\label{tab:process}
\end{table}
\normalsize
Note that it is possible to define a sequence of processes --- that is, a series of processes to be run in a single \ts{FeynMaster} run; this is perhaps more clear in the \t{Control.py} file itself, where the comments in the \textit{Processes section} show the line where the first process begins (\textit{process 1}) and the line where it ends (\textit{end of process 1}). To define a sequence of processes, the user must copy the lines of the first process, paste them after it, and edit them to define the second process, and repeat the same procedure for more processes. An example of a sequence of processes will be given in section \ref{sec:FullReno}.
For now, we must clarify the meaning of the variables of Table \ref{tab:process}.

\n \t{InParticles} and \t{OutParticles}, corresponding to the incoming and outgoing particles of the process, should contain only particles defined in \ts{QGRAF} style in the \ts{FeynMaster} model. Each particle should be written between primes, and different particles should be separated by commas (see example in Table \ref{tab:process}). Whenever both a particle and an antiparticle are defined in \t{InParticles} or \t{OutParticles}, the particle should always be written first, and the antiparticle after it. Tadpoles are obtained by selecting a single incoming particle and no outgoing particles (i.e., \t{OutParticles} = [\,]).

\n The \t{loops} variable should be set to a non-negative integer. Whatever the number of loops, \ts{FeynMaster} will always correctly generate the amplitudes for every diagram involved, although it is only prepared to properly draw and compute diagrams with number of loops inferior to 2. 

\n \t{Parsel} allows the specification of intermediate particles contributing to the process.\fn{It is similar to (and actually based on) the \t{iprop} option in \ts{QGRAF}.} It applies not only to particles in loops, but to all intermediate particles. Inside the outer squared brackets, specific selections (themselves defined by squared brackets) can be placed, in such a way that different selections should be separated by commas (see example in Table \ref{tab:process}, where we defined two specific selections). Each specific selection contains four arguments: the first should either be \t{`avoid'} or \t{`keep'}, the second should correspond to a particle of the model,\fn{Care should be taken not to select antiparticles, but only particles. This is because the propagator in \ts{FeynMaster} is defined through the particle, and not the antiparticle.} and the last two should be non-negative integer numbers such that the second is not smaller than the first. We illustrate how it works by considering the example in Table \ref{tab:process}: \t{[`avoid',`WP',`1',`3']} discards all the diagrams with number of `WP' propagators between 1 and 3, while \t{[`keep',`b',`1',`1']} keeps only diagrams with number of `b' propagators between 1 and 1 (i.e., exactly equal to 1).

\n \t{LoopTec} enables the user to choose between different loop techniques in \ts{FeynCalc}. There are four possibilities: 1 for \t{OneLoop}, 2 for \t{MyOneLoop}, 3 for \t{MyOneLoopMod}, 4 for \t{OneLoopTID}. The first possibility is the standard \t{OneLoop} included in \ts{FeynCalc}. This function has two problems. First, it is limited to the third power of the loop momenta in the numerator. We have implemented the case of four momenta for diagrams with four denominators using the conventions of \ts{LoopTools}; this corresponds to the second possibility, \t{MyOneLoop}, which we have tested with the light-by-light scattering in QED. The second problem with the standard \t{OneLoop} function is that, sometimes (e.g. for long lines with exterior fermions at the end), it conflicts with the \t{DiracSimplify} function.\fn{Hence, whenever there are external fermions, it is safer not to use the loop techniques 1 and 2.} One way to avoid such problem is to use the \t{TID} function included in \ts{FeynCalc}; this corresponds to the fourth possibility, \t{OneLoopTID}, which calls the same arguments as \t{OneLoop}. However, we have developed another algorithm for the decomposition of the numerator that is normally faster than the \t{TID} decomposition. This corresponds to the third possibility, \t{MyOneLoopMod}, which also includes the case of four momenta in the numerator.

\n \t{factor} is a number, written in \t{\ts{Mathematica}} style and between primes, that is to be factored out in the final calculations. Finally, \t{options} refers to \ts{QGRAF} options.\fn{See \ts{QGRAF} manual. Other examples besides `onepi' (for one particle irreducible diagrams only) are \t{`'} (no options) and \t{`notadpole'} (no tadpoles).}

\n Once the process (or sequence of processes) is specified, the last thing to edit before running \ts{FeynMaster} is the \textit{Selection section} of \t{Control.py}. It contains 9 logical variables (i.e., variables that can only be set either to True or to False), described in Table \ref{tab:sel}. 
\begin{table}[!h]%
\begin{normalsize}
\normalsize
\begin{center}
\begin{tabular}
%{@{}p{3cm} l @{}}
{@{\hspace{3mm}}>{\raggedright\arraybackslash}p{2.8cm}>{\raggedright\arraybackslash}p{7.2cm}@{\hspace{3mm}}}
\hlinewd{1.1pt}
Variable & Effect (when chosen as True) \\
\hline\\[-1.5mm]
\t{FRinterLogic} & establish an interface with \ts{FeynRules} \\[2.5mm] 
\t{RenoLogic} & perform renormalization \\[2.5mm]
\t{Draw} & draw and print the Feynman diagrams \\[2.5mm]
\t{Comp} & compute and print the final expressions \\[2.5mm]
\ \ \t{FinLogic} & print the final result of each diagram \\[2.5mm]
\ \ \t{DivLogic} & print the divergent part of each diagram \\[2.5mm]
\t{SumLogic} & compute and print the sum of the expressions\\[2.5mm]
\t{MoCoLogic} & apply momentum conservation \\[2.5mm]
\t{LoSpinors} & include spinors \\[2.5mm]
\hlinewd{1.1pt}
\end{tabular}
\end{center}
\vspace{-5mm}
\end{normalsize}
\caption{Logical variables of the \textit{Selection section} of \t{Control.py}. See text for details.}
\label{tab:sel}
\end{table}
\normalsize

\n Some remarks are in order, concerning the effect of these variables when set to True.

\n \t{FRinterLogic}, by establishing an interface with \ts{FeynRules}, performs several tasks. First, it runs \ts{FeynRules} (for the model selected in the initial variable \t{model}), prints the complete non-renormalized Feynman rules of the model in a PDF file and opens this file.\fn{\label{nt:1st}Note that the results are automatically written; this is especially challenging when it comes to (automatically) breaking the lines in a long equation. This challenge is in general surpassed with the \LaTeX\, \t{breqn} package, which is employed by \ts{FeynMaster}. However, \t{breqn} is not able to break a line whenever the point where the line is to be broken is surrounded by three or more parentheses; in those cases, unfortunately, the lines in the \ts{FeynMaster} PDF outputs simply go out of the screen. For documentation on the \t{breqn} package, cf.\url{https://www.ctan.org/pkg/breqn}.}
Second, it generates a \ts{QGRAF} model, a crucial element in the generation of Feynman diagrams. Third, it generates the complete non-renormalized Feynman rules in \ts{FeynCalc} style, which will play a decisive role in all the calculations. Finally, it generates a \t{\ts{Mathematica}} notebook specifically designed to run \ts{FeynRules} --- hereafter the \ts{FeynRules} notebook. This notebook is very useful in case the user wants to have control over the generation of Feynman rules, and is the subject of section \ref{sec:FR}.

\n Note that, even if all logical variables are set to False, \ts{FeynMaster} generates a \t{\ts{Mathematica}} notebook specifically designed to run \ts{FeynCalc} --- hereafter the \ts{FeynCalc} notebook. This notebook is very useful should the user want to have control over calculations, and is the subject of section \ref{sec:FC}. Moreover, once there is a \ts{QGRAF} model, \ts{FeynMaster} always runs \ts{QGRAF}, which writes in a symbolic form the total diagrams that contribute to the process at stake --- the same process which was specified through the variables in Table \ref{tab:process}. Then, \ts{FeynMaster} takes the \ts{QGRAF} output and writes the series of amplitudes for each diagram in a file that the \ts{FeynCalc} notebook shall have access to.

\n \t{RenoLogic} concerns the renormalization of the model. If \t{FRinterLogic} is set to True, \t{RenoLogic} prints the complete set of Feynman rules for the counterterms interactions in a PDF file and opens this file; moreover, it stores those interactions in a file which the \t{\ts{Mathematica}} notebook shall have access to. A second important feature of \t{RenoLogic} is described below, in the context of the \t{Comp} variable.

\n \t{Draw} takes the \ts{QGRAF} output, draws the Feynman diagrams in a \LaTeX \,file, prints them in a PDF file and opens this file. This operation is achieved with the help of \t{feynmf} \cite{feynmf}, a \LaTeX \,package to draw Feynman diagrams. Since the diagrams are written in a \LaTeX \,file, they can not only be edited by the user, but also directly copied to the \LaTeX \,file of the user's paper.\fn{As already suggested, \t{Draw} is at present only guaranteed to properly draw the diagrams up to 1-loop. Moreover, diagrams with more than two particles in the initial or final states, as well as some reducible diagrams, are also not warranted.}

\n \t{Comp} computes the final expressions using \ts{FeynCalc}, prints them in a PDF file and opens this file.\fn{By `final expressions' we mean the simplified analytical expressions for the diagrams written in terms of Passarino--Veltman integrals; more details on section \ref{sec:FC}. We inform
	that it is normal that the warning `\textit{front end is not available}' shows up when \t{Comp} is set to True. Finally, the limitation we alluded to in note \ref{nt:1st} applies here too.}
\t{Comp} is intrinsically related to the three logical variables that follow, which we now turn to. The first two of them --- \t{FinLogic} and \t{DivLogic} --- are, in fact, nothing but options for \t{Comp}, so that they only make sense if \t{Comp} is set to True:\fn{In case \t{Comp} is set to False, the logical value of such variables is irrelevant: they can either be set to True or to False.} \t{FinLogic} includes the (total) final analytical expression for each diagram in the PDF file printed by \t{Comp}; \t{DivLogic}, on the other hand, includes (only) the analytical expression for the UV divergent part of each diagram in the same PDF file.

\n {At this point, we should clarify the difference between UV divergences and infrared (IR) divergences. It is well known that, while the former are in general present in loop integrals, the latter can only show up when there is a massless particle running inside the loop (in which case the IR divergence comes from the integration region near $k^2=0$, with $k$ the loop momentum). In the present version of \ts{FeynMaster}, we restrict the treatment of divergences to the UV ones. Indeed, we assume that the IR divergences can be regulated by giving the massless particle a fake mass --- which one shall eventually be able to set to zero in physical processes, after considering real emission graphs. With that assumption, IR divergences will never show up explicitly (only implicitly through the fake mass). In the following, unless in potentially dubious statements, we will stop writing UV explicitly: it is assumed that, whenever we mention divergences, we shall be referring to UV divergences.}

\n \t{SumLogic} is an option for  both \t{Comp} and the \ts{FeynCalc} notebook: it computes the sum of the analytical expressions; if \t{Comp} is set to True, this sum is included in the generated PDF;\fn{More specifically: if \t{FinLogic} is True, \t{SumLogic} includes the sum of the total final expressions; if \t{DivLogic} is True, it includes the sum of the expressions for the divergent parts; if both are True, it includes both the sum of the total expressions and the sum of the expressions for the divergent parts.} whatever the logical value of \t{Comp}, setting \t{SumLogic} to True implies that, when the \ts{FeynCalc} notebook is run, the sum of the analytical expressions is calculated.

\n We now explain the effect of \t{RenoLogic} when \t{Comp} is set to True.\fn{In this case, momentum conservation should be enforced; that is to say, the variable \t{MoCoLogic} should be set to True. Note, moreover, that there is only a non trivial effect if there are counterterms interactions stored; given what we saw above, this means that \ts{FeynMaster} had to be run with both \t{RenoLogic} and \t{FRinterLogic} set to True for the model at stake.} In case the user defined a single process in the \textit{Processes section}, \t{RenoLogic} causes \ts{FeynMaster} to look for counterterms that might absorb the divergences of the process at stake, and to calculate those counterterms in $\overline{\text{MS}}$.\fn{That is, calculates them in such a way that the counterterms are precisely equal to the divergent part they absorb (except for the $\ln(4\pi)$ and the Euler-Mascheroni  constant $\gamma$, which are also absorbed in the $\overline{\text{MS}}$ scheme). By `calculating' we mean here writing the analytical expression.} Such counterterms are then printed in the PDF file produced by \t{Comp}, and stored in yet another file (\t{CTfin.m}, to be described below). In the case of a sequence of processes, the subsequently computed counterterms are added to \t{CTfin.m}; however, what is particularly special about the sequence is that, for a certain process of the sequence, \ts{FeynMaster} will compute the counterterms by making use of the counterterms already computed in the previous processes.\fn{This is, in fact, the major advantage of writing a series of processes in a single \ts{FeynMaster} run (as opposed to one process per run).} In the end of the run, \t{CTfin.m} contains all the counterterms that were computed (in $\overline{\text{MS}}$) to absorb the divergences of the processes of the sequence. In this way, and by choosing an appropriate sequence of processes, it is possible to automatically renormalize the whole model in $\overline{\text{MS}}$ with a single \ts{FeynMaster} run.

\n The last two logical variables of Table \ref{tab:sel} are \t{MoCoLogic} and \t{LoSpinors}. The former applies momentum conservation to the final expressions, while the latter includes fermion spinors in those expressions. These variables are relevant even when \t{Comp} is set to False, as shall be explained in section \ref{sec:FC}.

\subsection{Run}
\label{sec:Run}

\n If both the \ts{FeynRules} model and the \ts{FeynMaster} model are defined, and if the \t{Control.py} file is edited, everything is set. To run \ts{FeynMaster}, just run the \t{RUN-\ts{FeynMaster}} batch file inside directory \t{dirFM}.\fn{Care should be taken not to run \ts{FeynMaster} when the relevant notebooks are open. More specifically, if \t{FRinterLogic} is set to True, and if the \ts{FeynRules} notebook created for the process at stake already exists, this notebook cannot be open during the run; in the same way, if \t{Comp} is set to True, and if the \ts{FeynCalc} notebook designed for the process at stake already exists, such notebook cannot be open during the run.}

\subsection{Outputs}
\label{sec:Output}

\n Depending on the logical value of the variables in the \textit{Selection section} of \t{Control.py}, \ts{FeynMaster} can have different outputs. We now list the total set of outputs, assuming that all those variables are set to True.\fn{Actually, when \t{Comp} and \t{RenoLogic} are both True, \t{LoSpinors} should be False. This is irrelevant for what follows, since \t{LoSpinors} has no influence on the outputs as a whole.} First, in the directory where the \ts{FeynRules} model is, two files are generated: the \ts{FeynRules} notebook, \t{Notebook.nb}, and an input file for it, \t{PreControl.m}. Second, the \ts{QGRAF} model with the name corresponding to the variable \t{intname} is created in the directory \t{dirQmod}; besides, a file named \t{last-output} (with the last output from QGRAF) is created inside \t{dirQ}. Then, if it does not exist yet, a directory with the name corresponding to \t{intname} is created in the directory \t{dirMain}. Inside it, and if they do not exist yet, three directories are created, \t{Counterterms}, \t{FeynmanRules} and \t{Processes}, which we now describe.

\n \t{Counterterms} contains one folder, \t{TeXs-drawing}, and two files, \t{CTini.m} and \t{CTfin.m}. \t{TeXs-drawing} is where the PDF file with the complete set of Feynman rules for the counterterms interactions is stored, as well as the \LaTeX \, file that creates it. \t{CTini.m} is the file which the \ts{FeynCalc} notebook has access to and where the Feynman rules for the counterterms interactions are stored. \t{CTfin.m}, in turn, is the aforementioned file containing the counterterms that were computed (in $\overline{\text{MS}}$) to absorb the divergences of the processes of the sequence at stake.\fn{Everytime \ts{FeynMaster} is run, this file \t{CTfin.m} is generated anew, thus erasing any counterterms that might be contained in it.}

\n \t{FeynmanRules}, besides several files with the Feynman rules to be used in the \ts{FeynCalc} notebook, contains yet another \t{TeXs-drawing} folder, where the PDF file with the complete  set of Feynman rules for the
non-renormalized interactions is stored, as well as the \LaTeX \, file that creates it.

\n \t{Processes} contains a folder for each of the different processes studied. These folders are named with the index (in the sequence of processes) corresponding to the process at stake, as well as with a string containing the \ts{QGRAF} names of the incoming and the outgoing particles joined together. Inside each folder, there are two other folders, \t{TeXs-drawing} and \t{TeXs-expressions}, as well as three files, \t{Amplitudes.m}, \t{Helper.m} and the \ts{FeynCalc} notebook, \t{Notebook.nb}. In order: \t{TeXs-drawing} contains the PDF file with the printed Feynman diagrams, as well as the \LaTeX \,file that creates it; \t{TeXs-expressions} contains the PDF file with the printed expressions, as well as the \LaTeX \,file that creates it; \t{Amplitudes.m} contains the amplitudes for the diagrams (written in \ts{FeynCalc} style); \t{Helper.m} is an auxiliary file for the \ts{FeynCalc} notebook.

\n Finally, recall that, in case there is already a QGRAF model, \ts{FeynMaster} will run even if all variables of Table \ref{tab:sel} are set to False. This is relevant since it generates not only the QGRAF output (\t{last-output}), but also the folder (or folders) for the specific process (or processes) selected, containing the files described above.\fn{
%While QGRAF, when run on its own, does not write over the output file, when run inside \ts{FeynMaster} it \textit{does} write over the output file.\ref{sec:NB}
While the QGRAF output is not overwritten when QGRAF is run on its own, it is overwritten when QGRAF is run inside \ts{FeynMaster}.}

\subsection{The notebooks}
\label{sec:NB}

\n As previously mentioned, a major advantage of \ts{FeynMaster} is its hybrid character concerning automatization. Indeed, not only does it generate automatically the results, but it also allows the user to handle them. This is realized due to the automatic creation of the \ts{FeynRules} notebook and the \ts{FeynCalc} notebook. We now describe them in detail.

\subsubsection{The \ts{FeynRules} notebook}
\label{sec:FR}

\n We mentioned in section \ref{sec:Control} that, when \ts{FeynMaster} is run with the logical variable \t{FRinterLogic} set to True, the \ts{FeynRules} notebook \t{Notebook.nb} is automatically created in the directory \t{dirFRmod}. By running it,\fn{The run will generate several \ts{FeynMaster} internal files, among which is \t{built-model}, the \ts{QGRAF} model file.} the user can access the vertices for the different Lagrangean parts, according to Table \ref{tab:FRverts}.
\begin{table}[!h]%
\begin{normalsize}
\normalsize
\begin{center}
\begin{tabular}
%{@{}p{3cm} l @{}}
{@{\hspace{3mm}}>{\raggedright\arraybackslash}p{3.2cm}>{\raggedright\arraybackslash}p{3.8cm}@{\hspace{3mm}}}
\hlinewd{1.1pt}
Lagrangean part & vertices \\
\hline
\t{LGauge} & \t{vertsGauge} \\
\t{LFermions} & \t{vertsFermionsFlavor} \\
\t{LYukawa} & \t{vertsYukawa} \\
\t{LHiggs} & \t{vertsHiggs} \\
\t{LGhost} & \t{vertsGhosts} \\
\hlinewd{1.1pt}
\end{tabular}
\end{center}
%\end{footnotesize}
\vspace{-5mm}
%\end{large}
\end{normalsize}
\caption{Names of the different vertices according to the Lagrangean part (compare with Table \ref{tab:FRnames}).}
\label{tab:FRverts}
\end{table}
\normalsize
Besides the usual \ts{FeynRules} instructions, two useful functions --- \t{GetCT} and \t{MyTeXForm} --- are available.
\t{GetCT} is a function that, for a certain Lagrangean piece given as argument, yields the respective counterterms Lagrangean.
\t{MyTeXForm} is \ts{FeynMaster}'s version of \t{\ts{Mathematica}}'s \t{TeXForm}; it is a function that uses \t{\ts{Python}} (as well as inner \ts{FeynMaster} information concerning the \LaTeX \,form of the parameters of the model) to write expressions in a proper \LaTeX \,form.\fn{\t{MyTeXForm} prints the \LaTeX \,form of the expression at stake not only on the screen, but also in an external file named \t{MyTeXForm-last-output.tex} in the directory where the notebook lies.}

\subsubsection{The \ts{FeynCalc} notebook}
\label{sec:FC}

\n Whenever \ts{FeynMaster} is run, and independently of the logical values of the variables of Table \ref{tab:sel}, the \ts{FeynCalc} notebook is automatically created. This notebook, as already the \ts{FeynRules} one just described, is totally ready-to-use: the user does not have to define directories, nor import files, nor change conventions. Just by running the notebook, there is immediate access to a whole set of results: not only the totality of the results obtained should the \t{Comp} logical variable had been turned on (the expressions, the divergent parts, the counterterms, etc.), but also to more basic elements, such as the total Feynman rules for the model and amplitudes for the diagrams. Besides, since all these results are written in a \t{\ts{Mathematica}} notebook, the user has great control over them, as he or she can operate algebraically on them, or select part of them, or print them into files, etc. Moreover, since the \ts{FeynCalc} package is loaded, and since all the results are written in a \ts{FeynCalc}-readable style, the control at stake is even greater, for the user can apply all the useful tools of that package: operate on the Dirac algebra, perform contractions, solve loop integrals, etc.\fn{In the following, we assume the user to be familiar with \ts{FeynCalc}. For more informations, consult the \ts{FeynCalc} website: \url{https://feyncalc.github.io/}.} 

\n We now present some useful features introduced by \ts{FeynMaster} in the \ts{FeynCalc} notebook. We start with the variables related to the analytical expressions for the Feynman diagrams: see Table \ref{tab:FCvars}.
\begin{table}[!h]%
\begin{normalsize}
\normalsize
\begin{center}
\begin{tabular}
%{@{}p{3cm} l @{}}
{@{\hspace{3mm}}>{\raggedright\arraybackslash}p{1.8cm}>{\raggedright\arraybackslash}p{8.0cm}@{\hspace{3mm}}}
\hlinewd{1.1pt}
Variable & Meaning \\
\hline\\[-1.5mm]
\t{amp} & list with all the amplitudes \\[2.5mm]
\t{amp}\textit{i} & amplitude for diagram \textit{i} \\[2.5mm]
\t{ans} & list with all the modified amplitudes \\[2.5mm]
\t{ans}\textit{i} & modified amplitude for diagram \textit{i} \\[2.5mm]
\t{res} & list with all the final expressions \\[2.5mm]
\t{res}\textit{i} & final expression for diagram \textit{i} \\[2.5mm]
\t{resD} & list with all the expressions for the divergent parts \\[2.5mm]
\t{resD}\textit{i} & expression for the divergent part of diagram \textit{i} \\[2.5mm]
\t{restot} & sum of all the final expressions \\[2.5mm]
\t{resDtot} & sum of all the expressions for the divergent parts \\[2.5mm]
\hlinewd{1.1pt}
\end{tabular}
\end{center}
\vspace{-5mm}
\end{normalsize}
\caption{Useful variables concerning expressions for the diagrams. See text for details.}
\label{tab:FCvars}
\end{table}
\normalsize
Here, the differences between \t{amp}, \t{ans} and \t{res} should be highlighted. First, \t{amp} is a list that, for each element, contains the mere conjunction of the Feynman rules involved in the diagram at stake; \t{ans} is the result of the application of the loop technique in dimensional regularization to \t{amp} (in case \t{amp} is at 1-loop), or a mere simplification of \t{amp} (in case \t{amp} is at tree-level);
moreover, \t{ans} multiplies \t{amp} by $1/i$;
\t{res} takes \t{ans} and writes it in 4 dimensions --- including possible finite parts coming from this conversion\fn{As is well known, in the {dimensional regularization scheme}, the infinities are tamed by changing the dimensions of the integrals from 4 to $d$, in such a way that the divergences are regulated by the parameter $\epsilon = 4 - d$. When solving the integrals in terms of Passarino--Veltman integrals, the result will in general depend explicitly on the dimension $d$, as well as on the Passarino--Veltman integrals themselves --- which usually diverge, with divergence proportional to $1/\epsilon$. But since $d = 4-\epsilon$, there will in general be finite terms (order $\epsilon^0$) coming from the product between $d$ and the divergent parts in the Passarino--Veltman integrals. Hence, when converting the result back to 4 dimensions (since the final result is written in 4 dimensions), one cannot forget to include such terms. {Finally, recall that IR divergences will never show up explicitly if the potentially IR divergent integrals are tamed by giving the massless particle a fake mass.}\label{note:DimReg}} --- and factorizes the previously selected \t{factor}.
We should stress that, since \t{ans} includes a factor $1/i$, the final results (\t{res}) correspond to $\mathcal{M}$, and not to $i \mathcal{M}$.
Finally, the divergent parts are written in terms of the variable \t{div}, defined as:
\begin{equation*}
	\t{div} = \dfrac{1}{2} \left( \dfrac{2}{\epsilon} - \gamma + \ln 4\pi \right).
\end{equation*}
(Note that, in the PDF file with the printed expressions, we change the name \t{div} to $\omega_{\epsilon}$.)

\n Next, we consider useful functions to manipulate the results: see Table \ref{tab:FCfunc}.
\begin{table}[!h]%
\begin{normalsize}
\normalsize
\begin{center}
\begin{threeparttable}
\begin{tabular}
%{@{}p{3cm} l @{}}
{@{\hspace{3mm}}>{\raggedright\arraybackslash}p{2.5cm}>{\raggedright\arraybackslash}p{10.9cm}@{\hspace{3mm}}}
\hlinewd{1.1pt}
Function & Action \\
\hline\\[-1.5mm]
\t{MyTeXForm} & write expressions in a proper \LaTeX \,form  \\[2.5mm]
\t{MyPaVeReduce} & apply \ts{FeynCalc}'s \t{PaVeReduce} and convert to 4 dimensions\\[2.5mm]
\t{DecayWidth} & calculate the decay width\tnote{$\star$}\\[2.5mm]
\t{DiffXS} & calculate the differential cross section\tnote{$\dagger$}\\[2.5mm]
\t{FCtoLT} & convert expressions to \ts{LoopTools} \\[2.5mm]
\t{TrG5} & calculate the trace for expressions with $\gamma_5$\\[2.5mm]
\t{ChangeTo4} & change to 4 dimensions\\[2.5mm]
\t{GetDiv} & get the divergent part of an expression \\[2.5mm]
\t{GetFinite} & get the finite part of an expression in dimensional regularization\tnote{$\ddagger$}\\[2.5mm]
\t{MyOneLoop} & extends \t{OneLoop} to four powers of momenta in numerator \tnote{$\#$} \\[2.5mm]
\t{MyOneLoopMod} & alternative to \t{MyOneLoop} \tnote{$\#$}\\[2.5mm]
\hlinewd{1.1pt}
\end{tabular}
\begin{tablenotes}
{\small
\item[$\star$] Only applicable to processes with 1 incoming and 2 outgoing particles.
\item[$\dagger$] Only applicable to processes with 2 incoming and 2 outgoing particles.
\item[$\ddagger$] Cf. note \ref{note:DimReg}.
\item[$\#$] Cf. section~\ref{sec:Control}. 
}
\end{tablenotes}
\end{threeparttable}
\end{center}
\vspace{-5mm}
\end{normalsize}
\vs{-1mm}
\caption{Useful functions. See text for details.}
\label{tab:FCfunc}
\end{table}
\normalsize
Some clarifications are in order here.
\t{MyTeXForm} is the same function as the one described in section \ref{sec:FR}. 
\t{MyPaVeReduce} is \ts{FeynMaster}'s version of \ts{FeynCalc}'s \t{PaVeReduce}; it applies \t{PaVeReduce} and writes the result in 4 dimensions --- again, not without including possible finite parts coming from this conversion.
Concerning \t{DecayWidth} and \t{DiffXS}, note that while the former is written solely in terms of masses, the latter is written also in terms of the center of momentum energy \t{S} as well of the scattering angle \t{Theta}.
\t{FCtoLT} is the function that allows the numerical interface of \ts{FeynMaster};  when applied to an expression, it generates three \t{\ts{Fortran}} files in the directory where the notebook lies: \t{MainLT.F}, \t{FunctionLT.F} and \t{MyParameters.h}{; the} first one, \t{MainLT.F}, is the beginning of a main \t{\ts{Fortran}} program, which must be completed according to the user's will{;} \t{MainLT.F} calls the function \t{MyFunction}, which is the \ts{LoopTools} version of the expression \t{FCtoLT} was applied to, and which is written in the second \t{\ts{Fortran}} file, \t{FunctionLT.F}{; in} turn, \t{MyParameters.h} contains the numerical values associated to the different parameters in the \ts{FeynRules} model file. {Last of all (we postpone the discussion about \t{TrG5} to the next paragraph), \t{GetDiv} yields the UV divergent part of an expression; care should be taken if IR divergences are not regulated via a fake mass, for in that case, although they can show up as poles of the Passarino--Veltman functions, they will not be detected by \t{GetDiv}. Moreover, \t{GetDiv} only yields the UV divergent part of the Passarino--Veltman functions \cite{Denner:2005nn,tca} that \ts{FeynCalc} and \ts{FeynMaster} can handle --- namely, integrals whose power of the loop momenta in the numerator is at most 3 (except for the $D$ functions, where we extended \ts{FeynCalc} up to the fourth power of the loop momenta).}

\n {Let us now consider \t{TrG5}. Since \ts{FeynMaster} is prepared to compute divergent integrals --- and, more specifically, to compute them via dimensional regularization ---, it defines Dirac and Lorentz structures (like $g^{\mu\nu}$ or $\gamma^{\mu}$) in dimension $d$, not in dimension 4. However, the definition of $\gamma_5$ in dimension $d$ is not trivial, as chiral fermions are a property of four dimensions. In fact, the treatment of $\gamma_5$ in dimensional regularization is still an open problem (see e.g. refs. \cite{Akyeampong:1973xi,Chanowitz:1979zu,Barroso:1990ti,Kreimer:1993bh,Jegerlehner:2000dz,Greiner:2002ui,Zerf:2019ynn}). By default, \ts{FeynMaster} assumes the so-called naive dimensional regularization scheme~\cite{Jegerlehner:2000dz}, which takes the relation $\{\gamma_5,\gamma^{\mu}\}=0$ to be valid in dimension $d$. This naive approach is applied both when the loop diagram has external fermions, and when it has only inner fermions (forming a closed loop). In the second case, in order to calculate the corresponding trace, \ts{FeynMaster} uses \t{TrG5}. This function starts by separating the expression it applies to in two terms: one with $\gamma_5$, another without $\gamma_5$. It then computes the trace of the former in $4$ dimensions, while keeping the dimension of the latter in its default value $d$.}\fn{It is a matter of course that the calculation of the term with $\gamma_5$ in dimension 4 can only be an issue when the integral multiplying it is divergent. This is simply because one does not need to regularize an integral that is not divergent. In particular, there is no need to use dimensional regularization for a finite integral, so that all the calculations can be made in dimension 4. Note also that \ts{FeynCalc} already includes different schemes to handle $\gamma_5$, and is expected to improve the treatment of $\gamma_5$ in dimensional regularization in future versions.}

\n Finally, we present in Table \ref{tab:FCreno} some useful variables concerning renormalization. 
\begin{table}[!h]%
\begin{normalsize}
\normalsize
\begin{center}
\begin{tabular}
%{@{}p{3cm} l @{}}
{@{\hspace{3mm}}>{\raggedright\arraybackslash}p{2.5cm}>{\raggedright\arraybackslash}p{12.0cm}@{\hspace{3mm}}}
\hlinewd{1.1pt}
Function & Meaning \\
\hline\\[-1.5mm]
\t{CT}\textit{process} & expression containing all the counterterms involved in the \textit{process} at stake \\[2.5mm]
\t{PreResReno} & sum of the total divergent part and \t{CT}\textit{process} \\[2.5mm]
\t{CTfinlist} & list with all the counterterms computed so far in $\overline{\text{MS}}$\\[2.5mm]
\t{ResReno} & the same as \t{PreResReno}, but using counterterms previously stored in CTfinlist\\[2.5mm]
\t{PosResReno} & the same as \t{ResReno}, but using also the counterterms calculated for the process at stake\\[2.5mm]
\hlinewd{1.1pt}
\end{tabular}
\end{center}
\vspace{-5mm}
\end{normalsize}
\caption{Useful variables concerning renormalization. See text for details.}
\label{tab:FCreno}
\end{table}
\normalsize
Two notes should be added. First, in the \t{CT}\textit{process} variable, \textit{process} corresponds to the \ts{QGRAF} names of the incoming and the outgoing particles joined together (e.g., in the Standard Model, for the process $h \to Z\gamma$, \t{CT}\textit{process} is \t{CTHZA}). Second, \t{PosResReno} should always be zero, since in the $\overline{\text{MS}}$ scheme the divergents parts are exactly absorbed by the counterterms.

\n Some final comments on the \ts{FeynCalc} notebook. First, the indices of the particles are described in Table \ref{tab:FCconv},
\begin{table}[!h]%
\begin{normalsize}
\normalsize
\begin{center}
\begin{tabular}
%{@{}p{3cm} l @{}}
{@{\hspace{3mm}}>{}m{2.0cm}>{\raggedright\arraybackslash}m{1.5cm}>{\raggedright\arraybackslash}m{1.7cm}>{\raggedright\arraybackslash}m{1.3cm}>{\raggedright\arraybackslash}m{1.3cm}>{\raggedright\arraybackslash}m{1.7cm}>{\raggedright\arraybackslash}m{1.7cm}@{\hspace{3mm}}}
\hlinewd{1.1pt}
& \small \ts{QGRAF} index & \small Lorentz index in \ts{FeynCalc} & \small Lorentz index in \LaTeX \,& \small Color index in \LaTeX \,& \small Momentum index in \ts{FeynCalc} & \small Momentum index in \LaTeX \,\\
\hline\\[-7.5mm]
\multirow{2}{2.5cm}{\\[-3mm]Incoming particles} & \multirow{2}{2.5cm}{-1\\[1.5mm]-3} & \multirow{2}{2.5cm}{\t{-J1}\\[1.5mm]\t{-J3}} & \multirow{2}{2.5cm}{$\mu$\\[1.5mm]$\rho$} & 
\multirow{2}{2.5cm}{$a$\\[1.5mm]$c$} & 
\multirow{2}{2.5cm}{\t{p1}\\[1.5mm]\t{p2}} & \multirow{2}{0.5cm}{\\[-0.2mm]$p_1$\\[1.7mm]$p_2$}\\[8mm]
\multirow{2}{2.5cm}{\\[-3mm]Outgoing particles}  & \multirow{2}{2.5cm}{-2\\[1.5mm]-4} & \multirow{2}{2.5cm}{\t{-J2}\\[1.5mm]\t{-J4}} & \multirow{2}{2.5cm}{$\nu$\\[1.5mm]$\sigma$} &
\multirow{2}{2.5cm}{$b$\\[1.5mm]$d$} &
\multirow{2}{2.5cm}{\t{q1}\\[1.5mm]\t{q2}} & \multirow{2}{0.5cm}{\\[2.7mm]$q_1$\\[1.7mm]$q_2$}\\[13mm]
\hlinewd{1.1pt}
\end{tabular}
\end{center}
\vspace{-5mm}
\end{normalsize}
\caption{Particle indices for the \ts{FeynCalc} notebook. }
\label{tab:FCconv}
\end{table}
\normalsize
and momentum conservation can be applied through the replacement rule \t{MomCons}.\fn{Whenever there is one incoming particle and two outgoing particles, \t{MomCons} replaces p1 by the remaining momenta; in all the other cases, MomCons replaces q1 by the remaining momenta.} Second, the \t{Helper.m} file contains, among other definitions, both the \t{factor} (defined in the \textit{Selection section} of \t{Control.py}) as well as the \ts{FeynCalc} conventions (defined as \t{FCeqs} in the \ts{FeynMaster} model).
Third, even if \t{Comp} is set to False in \t{Control.py}, setting \t{MoCoLogic} and \t{LoSpinors} to True has consequences for the \ts{FeynCalc} notebook. Indeed, the former implies that, when the notebook is run, momentum conservation will be applied; as for the latter, it makes the amplitudes be written with fermion spinors.
Finally, the replacement rule \t{FCsimp} contains the simplifications for \ts{FeynCalc} (defined with the same name in the \ts{FeynMaster} model).

\section{Examples}
\label{sec:Examples}

\subsection{Creation and complete automatic renormalization of a toy model}
\label{sec:FullReno}

\n Here we exemplify how to create a model, on the one hand, and how to completely renormalize it, on the other. The model will be very simple: QED with an extra fermion. We first show how to create such a toy model.

\n Probably the simplest way to create any model whatsoever is to copy and modify an already existing model. Recall that, in order to define a model in \ts{FeynMaster}, we need to create two files: a \ts{FeynRules} model file and a \ts{FeynMaster} model file. We start with the former; given the similarity between our toy model and QED, we duplicate the directory \t{dirFRmod} corresponding to the QED \ts{FeynRules} model, and name the duplicate \t{QED2}. We get inside \t{QED2}, and change the name of the \ts{FeynRules} model file from \t{QED.fr} to \t{QED2.fr}. We then open \t{QED2.fr}, and edit the model in three steps:\fn{We neglect here minor details like the internal \ts{FeynRules} name, which can be edited in the variable \t{M\$ModelName}.} first, we modify the parameter list to
\vs{5mm}
\begin{small}
\begin{addmargin}[8mm]{0mm}
\begin{verbatim}
(***** Parameter list ******)
M$Parameters = {
  mf1 == {TeX -> Subscript[m,1]},
  mf2 == {TeX -> Subscript[m,2]},
  ee == {TeX -> e},
  xiA == {TeX -> Subscript[\[Xi],A]}  }
\end{verbatim}
\end{addmargin}
\end{small}
\vs{3mm}
Then, in the Particle classes list, we slightly modify what we had, and we add a second fermion:\fn{The fermions are defined both in terms of Weyl spinors (the \t{W} variables) and Dirac spinors (the \t{F} variables). It is certainly true that, in models with no parity violation (like the present one), there is no need to define the fermions in terms of Weyl spinors. Nevertheless, we consider them for illustrative purposes.}
\vs{5mm}
\begin{small}
\begin{addmargin}[8mm]{0mm}
\begin{verbatim}
(***** Particle classes list ******)
M$ClassesDescription = {
   W[1] == {
      ClassName -> psi1L,
      SelfConjugate -> False,
      QuantumNumbers -> {Q-> Q},
      Chirality -> Left},
   W[2] == {
      ClassName -> chi1R,
      SelfConjugate -> False,
      QuantumNumbers -> {Q-> Q},
      Chirality -> Right},
   F[1] == {
      ClassName -> f1,
      SelfConjugate -> False,
      QuantumNumbers -> {Q-> Q},
      Mass -> mf1,
      WeylComponents -> {psi1L, chi1R}},
   W[3] == {
      ClassName -> psi2L,
      SelfConjugate -> False,
      QuantumNumbers -> {Q-> Q},
      Chirality -> Left},
   W[4] == {
      ClassName -> chi2R,
      SelfConjugate -> False,
      QuantumNumbers -> {Q-> Q},
      Chirality -> Right},
   F[2] == {
      ClassName -> f2,
      SelfConjugate -> False,
      QuantumNumbers -> {Q-> Q},
      Mass -> mf2,      
      WeylComponents -> {psi2L, chi2R}},  
  V[1] == {
      ClassName -> A,
      Mass -> 0,
      SelfConjugate -> True}  }
\end{verbatim}
\end{addmargin}
\end{small}
\vs{3mm}
Finally, we modify the Lagrangean to include a second fermion:
\vs{3mm}
\begin{small}
\begin{addmargin}[8mm]{0mm}
\begin{verbatim}
LGauge := -1/4 FS[A, \[Mu], \[Nu]] FS[A, \[Mu], \[Nu]]
LFermions := I psi1Lbar.sibar[mu].del[psi1L, mu] + I chi1Rbar.si[mu].del[chi1R, mu] \
				 - mf1 (psi1Lbar.chi1R + chi1Rbar.psi1L) \
				 + ee psi1Lbar.sibar[mu].psi1L A[mu] + ee chi1Rbar.si[mu].chi1R A[mu] \
				+ I psi2Lbar.sibar[mu].del[psi2L, mu] + I chi2Rbar.si[mu].del[chi2R, mu] \
				 - mf2 (psi2Lbar.chi2R + chi2Rbar.psi2L) \
				 + ee psi2Lbar.sibar[mu].psi2L A[mu] + ee chi2Rbar.si[mu].chi2R A[mu]
LGF := -1/2/xiA del[A[mu], mu] del[A[nu], nu]				 
\end{verbatim}
\end{addmargin}
\end{small}
\vs{3mm}
This completes the \ts{FeynRules} model. We now move to the directory where the \ts{FeynMaster} models are; we duplicate the file \t{QED.py}, and we name the duplicate \t{QED2.py}. We open \t{QED2.py}, and start by changing \t{intname} to \t{`QED2'} and \t{extname} to \t{`QED with two fermions'}. We then edit the following variables according to:\fn{Recall that the QGRAF name of an antifermion should always be the capitalized version of the QGRAF name of the respective fermion. Moreover, in order to avoid internal conflicts in \ts{FeynMaster}, the \ts{FeynRules} particles names should never be equal to any QGRAF antiparticles names; so, for example, if we had defined the \ts{FeynRules} particles with names \t{F1} and \t{F2},  we should never define the QGRAF antiparticles names as \t{[`F1',`F2']}. Finally, since there is no need to include in \t{FRtoQ} neither the trivial replacement rules nor the antiparticles replacement rules, \t{FRtoQ} is empty in this case.}
\vs{5mm}
\begin{small}
\begin{addmargin}[8mm]{0mm}
\begin{Verbatim}[commandchars=\\\{\}]
(...)
	`fel' : [`f1',`f2],    \textcolor{mygray}{# - - - fermions}
	`anfel'  : [`F1',`F2'],    \textcolor{mygray}{# - - - antifermions}
(...)
	`LAfel' : [`f_1',`f_2'],
	`LAanfel'  : [`\textbackslash\textbackslash \!\!bar\{f\}_1',`\textbackslash\textbackslash \!\!bar\{f\}_2'],
(...)
	`FRtoQ' : `\{\}',
\end{Verbatim}
\end{addmargin}
\end{small}
\vs{5mm}
Finally, we consider the variables concerning renormalization. In QED, we renormalize the theory using the replacements
\bs 
\begin{align}
	A_{\mu} & \to A_{\mu} + \frac{1}{2} \delta Z_3 \, A_{\mu}, \\
	\psi_L & \to \psi_L + \frac{1}{2} \delta Z^f_L \, \psi_L, \\
	\chi_R & \to \chi_R + \frac{1}{2} \delta Z^f_R \, \chi_R, \\
	e & \to e + \delta e \hs{1mm} e, \\
	m_f & \to m_f + \delta m_f \hs{1mm},
\end{align}
\label{eq:QED}
\es 
\hs{-2.7mm} where $\psi_L$ and $\chi_R$ are the left and right components of the fermion, respectively. In \t{QED2}, we shall have exactly the same replacement rules, but with an extra fermion. Therefore, we write in \t{QED2.py}:
\vs{5mm}
\begin{small}
\begin{addmargin}[8mm]{0mm}
\begin{verbatim}
(...)
	`renconsnum' : [`\\[Delta]ee', `\\[Delta]Z3', `\\[Delta]ZfL1', `\\[Delta]ZfR1', 
                  `\\[Delta]mf1',`\\[Delta]ZfL2', `\\[Delta]ZfR2', `\\[Delta]mf2'],
	`LArenconsnum' : [`\\delta e',`\\delta Z_3',`\\delta Z^{f,L}_1',`\\delta Z^{f,R}_1',
                   `\\delta m_1',`\\delta Z^{f,L}_2',`\\delta Z^{f,R}_2',`\\delta m_2'],
(...)
	`renorrules' : [`A[\\[Nu]_] -> A[\\[Nu]] + (1/2) \\[Delta]Z3 A[\\[Nu]]', \
                `ee -> ee + \\[Delta]ee ee', \
                `psi1L -> psi1L + (1/2) \\[Delta]ZfL1 psi1L', \
                `chi1R -> chi1R + (1/2) \\[Delta]ZfR1 chi1R', \
                `psi1Lbar -> psi1Lbar + (1/2) \\[Delta]ZfL1 psi1Lbar', \
                `chi1Rbar -> chi1Rbar + (1/2) \\[Delta]ZfR1 chi1Rbar', \
                `mf1 -> mf1 + \\[Delta]mf1',
                `psi2L -> psi2L + (1/2) \\[Delta]ZfL2 psi2L', \
                `chi2R -> chi2R + (1/2) \\[Delta]ZfR2 chi2R', \
                `psi2Lbar -> psi2Lbar + (1/2) \\[Delta]ZfL2 psi2Lbar', \
                `chi2Rbar -> chi2Rbar + (1/2) \\[Delta]ZfR2 chi2Rbar', \
                `mf2 -> mf2 + \\[Delta]mf2']
}
\end{verbatim}
\end{addmargin}
\end{small}
\vs{3mm}
This completes the \ts{FeynMaster} model, and hence the total specification of \t{QED2}.

\n Now that the model is totally specified, we want to proceed to its complete automatic renormalization --- that is, to the determination of the complete set of counterterms (in the $\overline{\text{MS}}$ scheme).
To do so, we open \t{Control.py} (we assume that the user has already specified all the required directories); we start by choosing \t{model = `QED2'}. Then, we must choose a sequence of processes such that all the counterterms are computed. To do so, note that the total set of counterterms is:
\be
\delta Z_3, \quad \delta Z^{f,L}_1, \, \delta Z^{f,R}_1, \, \delta m_1, \quad \delta Z^{f,L}_2, \, \delta Z^{f,R}_2, \, \delta m_2, \quad \delta e.
\ee
However, from the renormalization of QED, we know that the first one, $\delta Z_3$, can be determined by the vacuum polarization of the photon; the following three, $\delta Z^{f,L}_1, \, \delta Z^{f,R}_1, \, \delta m_1$, can be determined by the self-energy of $f_1$; by the same token, $\delta Z^{f,L}_2, \, \delta Z^{f,R}_2, \, \delta m_2$ can be determined by the self-energy of $f_2$; finally, $\delta e$ can be determined by one of the vertices (either $f_1 \bar{f}_1 \gamma$ or $f_2 \bar{f}_2 \gamma$) at 1 loop. From all this, we conclude that the \textit{Processes section} in \t{Control.py} should be:\fn{Recall that, whenever a 1-loop process has external fermions, the variable \t{LoopTec} cannot be set equal to 1 or 2.}
\vs{5mm}
\begin{small}
\begin{addmargin}[8mm]{0mm}
\begin{Verbatim}[commandchars=\\\{\}]
def processes(): #:::::::::::::::::::::::::::::::: Processes section ::::::: (...)
\textcolor{mygray}{# - - - - - - (do not edit):}
	Seq=[]; Loo=[]; LooTec=[]; Fac=[]; Opt=[]; ParS=[]
\textcolor{mygray}{# - - - - - - Write the sequence of 1-loop processes you want to study, in QGRAF style:}
	\textcolor{mygray}{# - - - - - - process 1:}
	\textcolor{mygray}{# - - - - - - - - - - incoming particles, separated by (...)}
	InParticles = [`A']
	\textcolor{mygray}{# - - - - - - - - - - outgoing particles, separated by (...)}
	OutParticles = [`A']
	\textcolor{mygray}{# - - - - - - - - - - number of loops, without prime symbols (example: 1):}
	loops = 1
	\textcolor{mygray}{# - - - - - - - - - - selection of particles in the loop: keep or avoid an (...)}
	\textcolor{mygray}{# - - - - - - - - - - example: Parsel = [[`avoid',`H',`1',`3'] (...)}
	Parsel = []
	\textcolor{mygray}{# - - - - - - - - - - loop technique in FeynCalc (1 for OneLoop, 2 for (...)}
	LoopTec = 1
	\textcolor{mygray}{# - - - - - - - - - - quantity to factor out in the end of the computation (...)}
	factor = `1'
	\textcolor{mygray}{# - - - - - - - - - - QGRAF options (example: `onepi'); else, leave it `':}
	options = `onepi'
	\textcolor{mygray}{# - - - - - - (do not edit):}
	Sum = [InParticles,OutParticles]; Seq.append(Sum); 	(...)
	\textcolor{mygray}{# - - - - - - end of process 1}
	\textcolor{mygray}{# - - - - - - process 2:}
	\textcolor{mygray}{# - - - - - - - - - - incoming particles, separated by (...)}
	InParticles = [`f1']
	\textcolor{mygray}{# - - - - - - - - - - outgoing particles, separated by (...)}
	OutParticles = [`f1']
	\textcolor{mygray}{# - - - - - - - - - - number of loops, without prime symbols (example: 1):}
	loops = 1
	\textcolor{mygray}{# - - - - - - - - - - selection of particles in the loop: keep or avoid an (...)}
	\textcolor{mygray}{# - - - - - - - - - - example: Parsel = [[`avoid',`H',`1',`3'] (...)}
	Parsel = []
	\textcolor{mygray}{# - - - - - - - - - - loop technique in FeynCalc (1 for OneLoop, 2 for (...)}
	LoopTec = 3
	\textcolor{mygray}{# - - - - - - - - - - quantity to factor out in the end of the computation (...)}
	factor = `1'
	\textcolor{mygray}{# - - - - - - - - - - QGRAF options (example: `onepi'); else, leave it `':}
	options = `onepi'
	\textcolor{mygray}{# - - - - - - (do not edit):}
	Sum = [InParticles,OutParticles]; Seq.append(Sum); 	(...)
	\textcolor{mygray}{# - - - - - - end of process 2}
	\textcolor{mygray}{# - - - - - - process 3:}
	\textcolor{mygray}{# - - - - - - - - - - incoming particles, separated by (...)}
	InParticles = [`f2']
	\textcolor{mygray}{# - - - - - - - - - - outgoing particles, separated by (...)}
	OutParticles = [`f2']
	\textcolor{mygray}{# - - - - - - - - - - number of loops, without prime symbols (example: 1):}
	loops = 1
	\textcolor{mygray}{# - - - - - - - - - - selection of particles in the loop: keep or avoid an (...)}
	\textcolor{mygray}{# - - - - - - - - - - example: Parsel = [[`avoid',`H',`1',`3'] (...)}
	Parsel = []
	\textcolor{mygray}{# - - - - - - - - - - loop technique in FeynCalc (1 for OneLoop, 2 for (...)}
	LoopTec = 3
	\textcolor{mygray}{# - - - - - - - - - - quantity to factor out in the end of the computation (...)}
	factor = `1'
	\textcolor{mygray}{# - - - - - - - - - - QGRAF options (example: `onepi'); else, leave it `':}
	options = `onepi'
	\textcolor{mygray}{# - - - - - - (do not edit):}
	Sum = [InParticles,OutParticles]; Seq.append(Sum) (...)
	\textcolor{mygray}{# - - - - - - end of process 3}
	\textcolor{mygray}{# - - - - - - process 4:}
	\textcolor{mygray}{# - - - - - - - - - - incoming particles, separated by (...)}
	InParticles = [`A']
	\textcolor{mygray}{# - - - - - - - - - - outgoing particles, separated by (...)}
	OutParticles = [`f1',`F1']
	\textcolor{mygray}{# - - - - - - - - - - number of loops, without prime symbols (example: 1):}
	loops = 1
	\textcolor{mygray}{# - - - - - - - - - - selection of particles in the loop: keep or avoid an (...)}
	\textcolor{mygray}{# - - - - - - - - - - example: Parsel = [[`avoid',`H',`1',`3'] (...)}
	Parsel = []
	\textcolor{mygray}{# - - - - - - - - - - loop technique in FeynCalc (1 for OneLoop, 2 for (...)}
	LoopTec = 3
	\textcolor{mygray}{# - - - - - - - - - - quantity to factor out in the end of the computation (...)}
	factor = `1'
	\textcolor{mygray}{# - - - - - - - - - - QGRAF options (example: `onepi'); else, leave it `':}
	options = `onepi'
	\textcolor{mygray}{# - - - - - - (do not edit):}
	Sum = [InParticles,OutParticles]; Seq.append(Sum) (...)
	\textcolor{mygray}{# - - - - - - end of process 4}
\textcolor{mygray}{# - - - - - - for more processes, just replicate the above process}
	return (Seq,Loo,ParS,LooTec,Fac,Opt)
\end{Verbatim}
\end{addmargin}
\end{small}
\vs{3mm}

\n Finally, concerning the variables of the \textit{Selection section} of \t{Control.py}, we set them all to True, except \t{LoSpinors}, which we set to False. This being done, everything is set. We then move to \t{dirFM}, and run batch the file \t{RUN-\ts{FeynMaster}}. In total, 10 PDF files are automatically and subsequentially generated and opened: one for the non-renormalizable Feynman rules, another one for the counterterms Feynman rules, and two files per process --- one with the Feynman diagrams, another with the respective expressions. In the last file for the expressions, we read ``\textit{This completes the renormalization of the model}'', and the list of the full set of counterterms is presented.

\subsection{$h \to \gamma\gamma$ in the Standard Model}
\label{sec:HAA}

\n In this example, we use the $h \to \gamma\gamma$ in the Standard Model as an illustration of several features of \ts{FeynMaster}.
We shall use Standard Model model files --- both the \ts{FeynRules} one and the \ts{FeynMaster} one --- that come with \ts{FeynMaster}, and we assume once again that the user has already specified all the required directories in \t{Control.py}. As for the remaining variables in this file, we set them in accordance with Table \ref{tab:FCexample}.\fn{\label{note:TrueFalse}
	{A \t{Control.py} file adapted to this example was included inside the folder containing the \ts{FeynRules} model for the Standard Model, under the name \t{Control-Example-5.2.py}.}
	We are setting \t{FRinterLogic} to True, which only needs to be done in case it was not yet done before. Actually, generating all Feynman rules for both the non-renormalizable interactions and the counterterms in the Standard Model may take some minutes. Therefore, if we have already performed that operation, we can save time by setting \t{FRinterLogic} and \t{RenoLogic} to False.}
\begin{table}[!h]%
\begin{normalsize}
\normalsize
\begin{center}
\begin{tabular}{@{\hspace{3mm}}>{\raggedright\arraybackslash}p{3cm}>{\raggedright\arraybackslash}p{3.0cm}@{\hspace{3mm}}}
\hlinewd{1.1pt}
Variable & Value \\
\hline\\[-1.5mm]
\t{model} & \t{`StandardModel'} \\[6mm]
\t{InParticles} & \t{[`H']} \\[2.5mm]
\t{InParticles} & \t{[`A',`A']} \\[2.5mm]
\t{loops} & \t{1} \\[2.5mm]
\t{Parsel} & \t{[]} \\[2.5mm]
\t{LoopTec} & \t{1} \\[2.5mm]
\t{factor} & \t{`1'} \\[2.5mm]
\t{options} & \t{`onepi'} \\[6mm]
\t{FRinterLogic} & True \\[2.5mm]
\t{RenoLogic} & True \\[2.5mm]
\t{Draw} & True \\[2.5mm]
\t{Comp} & False \\[2.5mm]
\t{FinLogic} & False \\[2.5mm]
\t{DivLogic} & False \\[2.5mm]
\t{SumLogic} & True \\[2.5mm]
\t{MoCoLogic} & False \\[2.5mm]
\t{LoSpinors} & False \\[2.5mm]
\hlinewd{1.1pt}
\end{tabular}
\end{center}
\vspace{-5mm}
\end{normalsize}
\caption{Values of the variables of \t{Control.py} for the example in section \ref{sec:HAA}; cf. note \ref{note:TrueFalse}.}
\label{tab:FCexample}
\end{table}
\normalsize
We then go to the directory \t{dirFM}, and we run the batch file \t{RUN-FeynMaster}. In total, 3 PDF files will automatically be generated and opened: one for the non-renormalizable Feynman rules, another one for the counterterms Feynman rules, and a third one for the Feynman diagrams. We move to the directory \t{1-HAA} inside \t{dirPro} (meanwhile generated) and open the \ts{FeynCalc} notebook \t{Notebook.nb}. We then run the \t{Notebook.nb}, after which we are ready to test some relevant features.

\subsubsection{Notebook access to Feynman rules}

\n First, we want to gain some intuition on how the notebook has access to the Feynman rules of the Standard Model and to the amplitudes of the $h \to \gamma\gamma$ decay. We write
\vs{1mm}
\begin{small}
\begin{addmargin}[8mm]{0mm}
\begin{Verbatim}[commandchars=\\\{\}]
\textcolor{mygray}{In[14]:=} amp1
\end{Verbatim}
\end{addmargin}
\end{small}
\vs{-2mm}
which should yield the expression:
\bd
\dfrac{2 \, e^3 \, {m_W} \, g^{-J2-J4}}{s_w \left(k_1^2 - {{m_W}}^2\right) \left((p_1-k_1)^2 - {{m_W}}^2\right)} \,  \,  - \dfrac{2 \, D \, e^3 \, {m_W} \, g^{-J2-J4}}{s_w \left(k_1^2 - {{m_W}}^2\right) \left((p_1-k_1)^2 - {{m_W}}^2\right)}.
\ed
This is the amplitude for the first Feynman diagram, where $D$ represents the dimension. Now, where does the notebook get this information from? To answer the question, we open \t{Amplitudes.m} inside \t{1-HAA}. If we check the first line, we realize that \t{amp1} is essentially a product of Feynman rules such as \t{propWP[...]} and \t{vrtxAAWPWM[...]}.\fn{Diagrams 7 to 24 have a factor \t{Nc}, which corresponds to the color number. This factor shall be present whenever there are fermions present.} These rules are defined either in \t{Feynman-Rules-Main.m} inside \t{dirFey}, or in one of the other files present in that directory. Although they have been automatically generated, they can always be edited for particular purposes.

\subsubsection{Finiteness and gauge invariance}
\label{sec:FT}

\n Next, we use some of the features described in section \ref{sec:FC} to test two important properties of $h \to \gamma\gamma$: finiteness and gauge invariance. We start with the former; by writing
\vs{1mm}
\begin{small}
\begin{addmargin}[8mm]{0mm}
\begin{Verbatim}[commandchars=\\\{\}]
\textcolor{mygray}{In[15]:=} resD
\end{Verbatim}
\end{addmargin}
\end{small}
\vs{-1mm}
we obtain the list with all the expressions for the divergents parts. It is a non-trivial list: although some of its elements are zero, most of them are not. However, when we sum the whole list, we find:
\vs{1mm}
\begin{small}
\begin{addmargin}[8mm]{0mm}
\begin{Verbatim}[commandchars=\\\{\}]
\textcolor{mygray}{In[16]:=} resDtot
\end{Verbatim}
\vs{-3mm}
\begin{Verbatim}[commandchars=\\\{\}]
\textcolor{mygray}{Out[16]=} 0
\end{Verbatim}
\end{addmargin}
\end{small}
\vs{-2mm}
so that the process as a whole is finite, as expected for this decay mode.

\n Let us now check gauge invariance. First of all, note that the total amplitude $M$ for $h \to \gamma\gamma$ can be written as
\be
M = \epsilon_1^{\nu} \epsilon_2^{\sigma} \, M_{\nu\sigma},
\ee
where we are just factoring out the polarization vectors $\epsilon_1^{\nu}$ and $\epsilon_2^{\sigma}$ of the two photons. Then, it is easy to show that gauge invariance forces $M^{\nu\sigma}$ to have the form
\be
M^{\nu\sigma} = ( g^{\nu\sigma} q_1 . q_2 - q_1^{\sigma}  q_2^{\nu}) F,
\label{eq:invgaugefinal}
\ee
where $q_1$ and $q_2$ are the 4-momenta of the two photons, and $F$ is a scalar function of the momenta and the masses. In other words, it is a consequence of gauge invariance that, in the total process, the coefficient of $g^{\nu\sigma} q_1 . q_2$ must be exactly opposite to that of $q_1^{\sigma}  q_2^{\nu}$. To test this, we define some replacement rules:
\begin{small}
\vs{1mm}
\begin{addmargin}[8mm]{0mm}
\begin{Verbatim}[commandchars=\\\{\}]
\hs{12mm} \textcolor{uglyblue}{(* momentum conservation in scalar products and four-vectors *)}
\end{Verbatim}
\vs{-3mm}
\begin{Verbatim}[commandchars=\\\{\}]
\textcolor{mygray}{In[17]:=} dist = \{SP[p1, x_] -> SP[q1, x] + SP[q2, x], FV[p1, x_] -> FV[q1, x] + FV[q2, x]\};
\end{Verbatim}
\vs{-3mm}
\begin{Verbatim}[commandchars=\\\{\}]
\hs{12mm} \textcolor{uglyblue}{(* external particles on-shell *)}
\end{Verbatim}
\vs{-3mm}
\begin{Verbatim}[commandchars=\\\{\}]
\textcolor{mygray}{In[18]:=} onshell = \{SP[q1, q1] -> 0, SP[q2, q2] -> 0, SP[p1, p1] -> mH^2\};
\end{Verbatim}
\vs{-3mm}
\begin{Verbatim}[commandchars=\\\{\}]
\hs{12mm} \textcolor{uglyblue}{(* kinematics *)}
\end{Verbatim}
\vs{-3mm}
\begin{Verbatim}[commandchars=\\\{\}]
\textcolor{mygray}{In[19:=} kin = \{SP[q1, q2] -> MH^2/2, SP[p1, q1] -> MH^2/2, SP[p1, q2] -> MH^2/2\};
\end{Verbatim}
\vs{-3mm}
\begin{Verbatim}[commandchars=\\\{\}]
\hs{12mm} \textcolor{uglyblue}{(* transversality of the external photons polarizations *)}
\end{Verbatim}
\vs{-3mm}
\begin{Verbatim}[commandchars=\\\{\}]
\textcolor{mygray}{In[20]:=} transv = \{FV[q1, -J2] -> 0, FV[q2, -J4] -> 0\};
\end{Verbatim}
\end{addmargin}
\vs{1mm}
\end{small}
which we use to define a new \t{res} list:
\vs{3mm}
\begin{small}
\begin{addmargin}[8mm]{0mm}
\begin{Verbatim}[commandchars=\\\{\}]
\textcolor{mygray}{In[21]:=} resnew = (res /. dist /. onshell /. kin /. transv) // Simplify;
\end{Verbatim}
\end{addmargin}
\end{small}
\vs{1mm}
Finally, we write the coefficients of $g^{\nu\sigma} q_1 . q_2$ and $q_1^{\sigma}  q_2^{\nu}$ as
\vs{3mm}
\begin{small}
\begin{addmargin}[8mm]{0mm}
\begin{Verbatim}[commandchars=\\\{\}]
\textcolor{mygray}{In[22]:=} resnewT = (Coefficient[resnew, MT[-J2, -J4]] // MyPaVeReduce) 
\hs{12mm}  /(MH^2/2) // Simplify // FCE;
\end{Verbatim}
\vs{-3mm}
\begin{Verbatim}[commandchars=\\\{\}]
\textcolor{mygray}{In[23]:=} resnewL = (Coefficient[resnew, FV[q1, -J4]*FV[q2,-J2]] // MyPaVeReduce)
\hs{12mm}  // Simplify // FCE;
\end{Verbatim}
\end{addmargin}
\end{small}
\vs{1mm}
respectively, to conclude that
\vs{3mm}
\begin{small}
\begin{addmargin}[8mm]{0mm}
\begin{Verbatim}[commandchars=\\\{\}]
\textcolor{mygray}{In[24]:=} Total[resnewT] + Total[resnewL] // Simplify
\end{Verbatim}
\vs{-3mm}
\begin{Verbatim}[commandchars=\\\{\}]
\textcolor{mygray}{Out[24]:=} 0
\end{Verbatim}
\end{addmargin}
\end{small}
in accordance with gauge invariance. For what follows, it is convenient to save the expressions for the total transverse and longitudinal part. We write
\vs{3mm}
\begin{small}
	\begin{addmargin}[8mm]{0mm}
		\begin{Verbatim}[commandchars=\\\{\}]
		\textcolor{mygray}{In[25]:=} FT = Total[resnewT] // Simplify;
		\end{Verbatim}
		\vs{-3mm}
		\begin{Verbatim}[commandchars=\\\{\}]
		\textcolor{mygray}{In[26]:=} FL = Total[resnewL] // Simplify;
		\end{Verbatim}
	\end{addmargin}
	\end{small}

\subsubsection{\t{MyTeXForm}}

\n We now illustrate how to use \t{MyTeXForm} inside the \ts{FeynCalc} notebook. Suppose we want to write the sum of final results for the diagrams with quartic vertices (diagrams 1 to 6)  in a \LaTeX \, document. We define the variable \t{toprint1} as
\vs{3mm}
\begin{small}
\begin{addmargin}[8mm]{0mm}
\begin{Verbatim}[commandchars=\\\{\}]
\textcolor{mygray}{In[27]:=} toprint1 = Sum[res[[i]], {i, 1, 6}] // Simplify
\end{Verbatim}
\end{addmargin}
\end{small}
after which we write
\vs{3mm}
\begin{small}
\begin{addmargin}[8mm]{0mm}
\begin{Verbatim}[commandchars=\\\{\}]
\textcolor{mygray}{In[28]:=} toprint1 // MyTeXForm
\end{Verbatim}
\end{addmargin}
\end{small}
If we now copy the outcome as plain text and paste it in a \LaTeX \, document like the present one, we get:
\bd
- \left( e^3 \,  \left(  \left( {m_h}^2 + 6 \, {m_W}^2 \right)  \, B_0\left(p_1^2, {m_W}^2, {m_W}^2\right) + {m_W}^2 \,  \left( -4 + B_0\left(q_1^2, {m_W}^2, {m_W}^2\right) + B_0\left(q_2^2, {m_W}^2, {m_W}^2\right) \right)  \right)  \, g^{\nu \sigma} \right) / \left( 16 \, {m_W} \, \pi^2 \, {s_w} \right).
\ed
Note that we didn't need to break the line manually in the \LaTeX \, equation. This is because we are using the \t{breqn} package, which automatically breaks lines in equations.\fn{For documentation, cf. \url{https://www.ctan.org/pkg/breqn} . Recall that the line breaking does not work when the point where the line is to be broken is involved in three or more parentheses.}

\subsubsection{\ts{LoopTools} interface}

\n We mentioned in the Introduction that \FM \, includes an interface with \LT.
We now want to show how it works in the context of $h \to \gamma\gamma$. Suppose we want to plot the decay width as a function of the Higgs mass; we could start by computing the total $h \to \gamma\gamma$ decay width:
\vs{3mm}
\begin{small}
\begin{addmargin}[8mm]{0mm}
\begin{Verbatim}[commandchars=\\\{\}]
\textcolor{mygray}{In[27]:=} X0 = restot // DecayWidth
\end{Verbatim}
\end{addmargin}
\end{small}
However, although this works, it takes a long time and produces large expressions. It is simpler to use the amplitude in the generic form of Eq.~\ref{eq:invgaugefinal}, that is,
\vs{3mm}
\begin{small}
\begin{addmargin}[8mm]{0mm}
\begin{Verbatim}[commandchars=\\\{\}]
\textcolor{mygray}{In[27]:=} X0 = F (MT[-J2, -J4] MH^2/2 -  FV[q1, -J4] FV[q2, -J2]) // DecayWidth 
\end{Verbatim}
\end{addmargin}
\end{small}
Then, we write
\vs{3mm}
\begin{small}
\begin{addmargin}[8mm]{0mm}
\begin{Verbatim}[commandchars=\\\{\}]
\textcolor{mygray}{In[28]:=} (X0 /. F -> Abs[FT] // Simplify) // FCtoLT
\end{Verbatim}
\end{addmargin}
\end{small}
where we replaced the generic variable \t{F} by the absolute value of the total transverse part, \t{FT}, defined above. 
As explained in section \ref{sec:FC}, the command \t{FCtoLT} generates three files: \t{MainLT.F}, \t{FunctionLT.F} and \t{MyParameters.h}. We open \t{MainLT.F}, and immediatly after the comments
\vs{3mm}
\begin{small}
\begin{addmargin}[8mm]{0mm}
\begin{Verbatim}[commandchars=\\\{\}]
\textcolor{mygray}{* Write now the rest of the program}
\end{Verbatim}
\end{addmargin}
\end{small}
we write\footnote{The parameters loaded from the file \t{MyParameter.h} cannot be changed inside the \ts{Fortran} program (\t{MainLT.F}). Hence, since we define the parameter \t{MH} as the Higgs boson mass, we name \t{xMH} the variable we are using to make the plot; in doing so, we must be careful to replace \t{MH} for \t{xMH} in the arguments of \t{MyFunction} inside the loop.}
\vs{1mm}
\begin{small}
\begin{addmargin}[8mm]{0mm}
\begin{Verbatim}[commandchars=\\\{\}]
xMH=38d0
do i=1,162
  xMH=xMH+1d0
  write(50,98)xMH,MyFunction(..., xMH, ...)
enddo
\end{Verbatim}
\end{addmargin}
\end{small}
where \t{50} and \t{98} represent the output file and the impression format, respectively. We are varying the Higgs mass from 38 GeV to 200 GeV in steps of 1 GeV. The result is presented in Fig.~{\ref{fig:HAA}.
\begin{figure}[htb]
	\centering
	\includegraphics[width=0.5\textwidth]{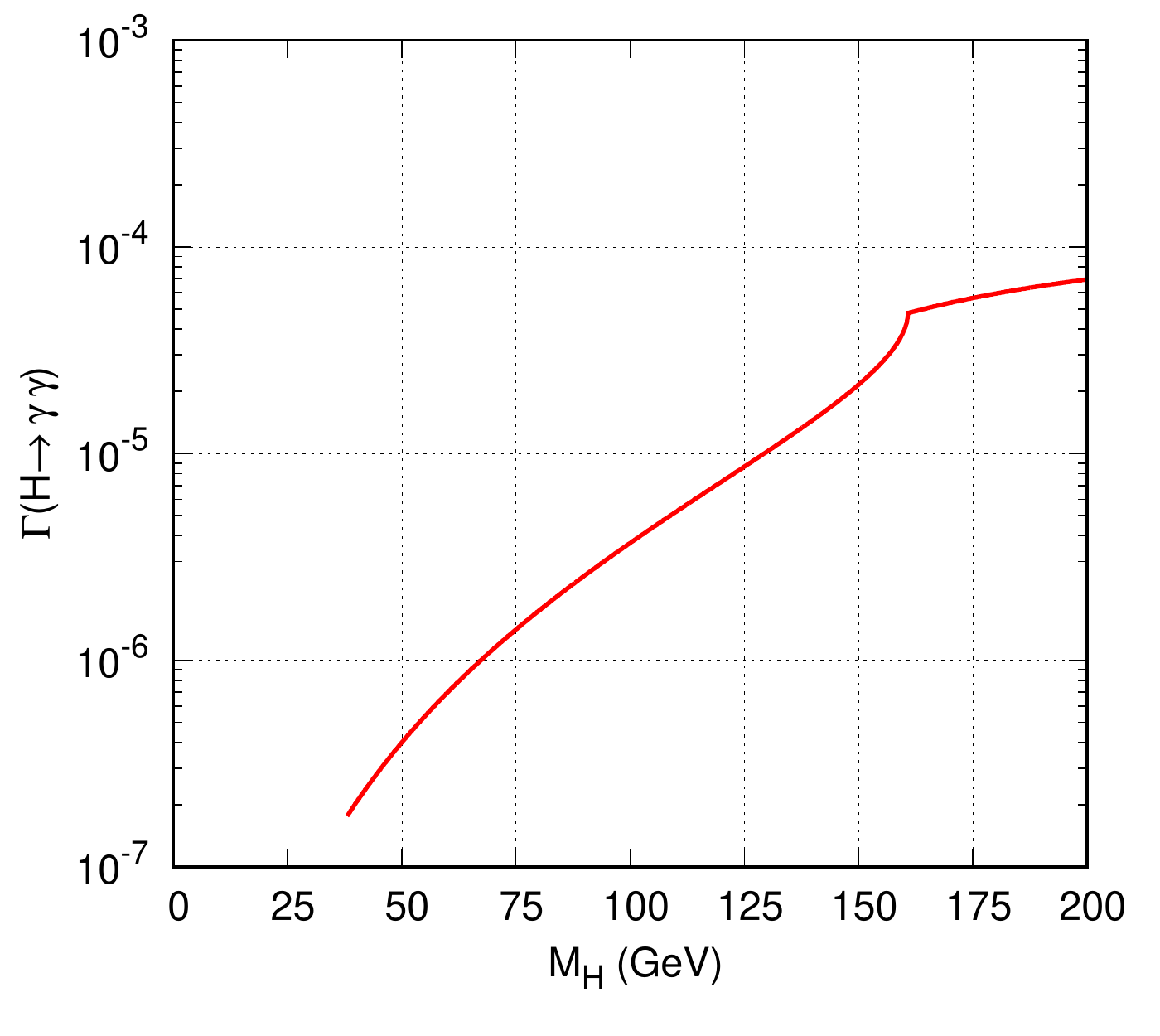}
	\caption{Width of the process $h \to \gamma \gamma$ as a function of the Higgs boson mass.}
	\label{fig:HAA}
\end{figure}	

\subsubsection{Edition of Feynman diagrams}

\n Finally, we briefly explain how to edit the Feynman diagrams. Recall that they were written in a \LaTeX \, file inside the \t{TeXs-drawing} folder, which in turn is located in the \t{dirPro} directory. If we open the file (\t{1-output-Drawing-HAA.tex}), we find the diagrams written according to the \t{feynmf} package. Let us consider the first diagram. The original code produces the original diagram:

\vs{3mm}
\hspace{7mm} 
\begin{minipage}[h]{.80\textwidth}
\begin{small}
\begin{verbatim}
(...)
\fmflabel{$\gamma$}{...} 
\fmflabel{$\gamma$}{...} 
\fmf{dashes,tension=3}{...} 
\fmf{photon,tension=3}{...} 
\fmf{photon,tension=3}{...} 
\fmf{photon,label=$W^{+}$,right=1}{...} 
\fmf{photon,label=$W^{+}$,right=1}{...} 
(...)
\end{verbatim}
\end{small}
\end{minipage}
\hspace{-45mm} 
\begin{minipage}[h]{.40\textwidth}
		\begin{picture}(0,80)
		\put(-30,-12){\includegraphics[width=6cm]{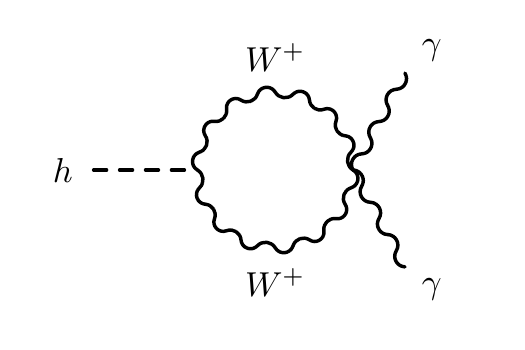}}
		\end{picture}
\end{minipage}
\vs{1mm}

\n However, we can modify the code in order to change the aspect of the diagram. In particular, we can change the labels, the tensions and the curvatures.\fn{The tensions represent the strenght of the lines: the larger the tension, the tighter the line will be. The default tension is 1. The curvature is represented by the variable \t{right}. Note that tensions, labels and curvatures are just a few examples of variables that can be changed to generate a different diagram. For more informations, please consult the \t{feynmf} manual.} We give two examples:

\vs{3mm}
\hspace{7mm} 
\begin{minipage}[h]{.80\textwidth}
\begin{small}
\begin{verbatim}
(...)
\fmflabel{$\gamma$}{...} 
\fmflabel{$\gamma$}{...} 
\fmf{dashes,tension=1}{...} 
\fmf{photon,tension=3}{...} 
\fmf{photon,tension=3}{...} 
\fmf{photon,label=$W^{-}$,right=1}{...} 
\fmf{photon,label=$W^{-}$,right=1}{...} 
(...)
\end{verbatim}
\end{small}
\end{minipage}
\hspace{-45mm} 
\begin{minipage}[h]{.40\textwidth}
		\begin{picture}(0,80)
		\put(-30,-12){ \includegraphics[width=6cm]{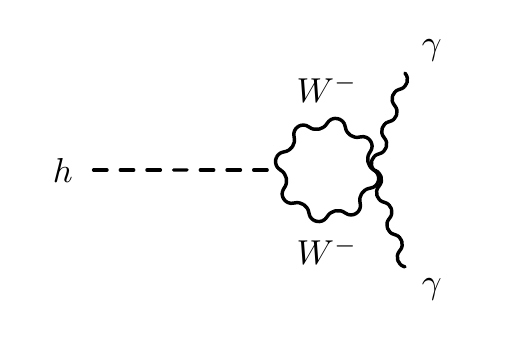}}
		\end{picture}
\end{minipage}
\vs{8mm}
\vs{3mm}
\hspace{7mm} 
\begin{minipage}[h]{.80\textwidth}
\begin{small}
\begin{verbatim}
(...)
\fmflabel{$\gamma_1$}{...} 
\fmflabel{$\gamma_2$}{...} 
\fmf{dashes,tension=5}{...} 
\fmf{photon,tension=3}{...} 
\fmf{photon,tension=3}{...} 
\fmf{photon,label=$W^{-}$,right=0.7}{...} 
\fmf{photon,label=$W^{-}$,right=0.7}{...} 
(...)
\end{verbatim}
\end{small}
\end{minipage}
\hspace{-45mm} 
\begin{minipage}[h]{.40\textwidth}
		\begin{picture}(0,80)
		\put(-20,-12){\includegraphics[width=6cm]{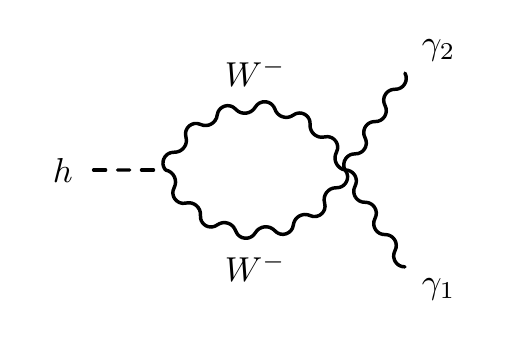}}
		\end{picture}
\end{minipage}
\vs{3mm}

\subsection{QED Ward identity}
\label{sec:WI}

\n In the previous example, we showed how to use \ts{FeynMaster} to manipulate the results of a single process. Here, we illustrate how it can also be used to combine information of several processes. For that purpose, we consider a simple task: prove the QED Ward identity.

\n It is easy to show that the Ward identity at 1-loop order in QED can be written as:
\be
p_1^{\nu} \, \Gamma_{\nu}(p_1,p_2,p_3) = e \left( \vphantom{\dfrac{A^B}{A^B}} \Sigma(p_2) - \Sigma(p_3) \right),
\label{eq:provanda}
\ee
where
\be
\Gamma_{\nu}(p_1,p_2,p_3) =
%\hs{8mm}
%\begin{minipage}{0.35\textwidth}
%\begin{fmffile}{QEDB11}
%\begin{fmfgraph*}(100,100) 
%\fmfset{arrow_len}{3mm} 
%\fmfset{arrow_ang}{20} 
%\fmfleft{nJ1}  
%\fmfright{nJ2,nJ4} 
%\fmf{photon,label=$p_1$,tension=4}{nJ1,J2J3nJ1} 
%\fmf{fermion,label=$p_2$,label.side=left,tension=4}{nJ2J1J5,nJ2} 
%\fmf{fermion,label=$p_3$,label.side=left,tension=4}{nJ4,J4nJ4J6} 
%\fmf{fermion,tension=1,label.dist=3thick}{J2J3nJ1,nJ2J1J5} 
%\fmf{fermion,tension=1,label.dist=3thick}{J4nJ4J6,J2J3nJ1} 
%\fmf{photon,label.side=left,tension=1,label.dist=3thick}{J4nJ4J6,nJ2J1J5} 
%\end{fmfgraph*}
%\end{fmffile}
%\end{minipage}
%\hs{-17mm},
%%
%\hs{15mm}
%%
\hs{8mm}
\begin{minipage}{0.35\textwidth}
		\begin{picture}(0,80)
		\put(-30,-9){\includegraphics[width=4cm]{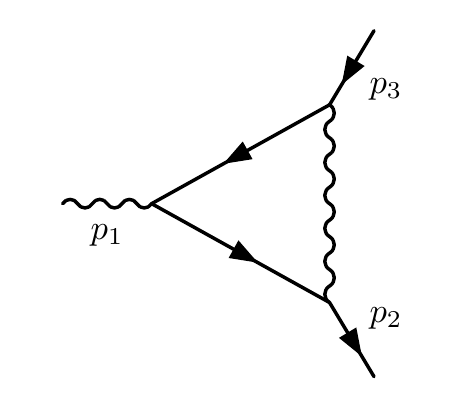}}
		\end{picture}
\end{minipage}
\hs{-17mm},
\hs{15mm}
\Sigma(p_i) = 
\begin{minipage}{0.35\textwidth}
		\begin{picture}(0,80)
		\put(-3,8){\includegraphics[width=4cm]{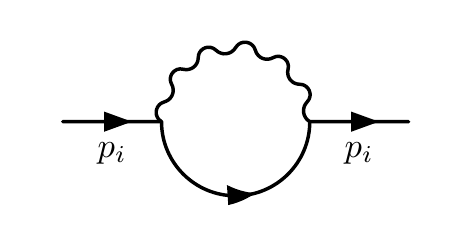}}
		\end{picture}
\end{minipage}
\hs{-15mm},
\label{eq:pic}
\ee
and where the momenta $p_1$ and $p_3$ are incoming, while $p_2$ is outgoing.
\begin{table}[!h]%
\begin{normalsize}
\normalsize
\begin{center}
\begin{tabular}{@{\hspace{6mm}}>{\raggedright\arraybackslash}p{3.2cm}>{\raggedright\arraybackslash}p{2.0cm}@{\hspace{6mm}}}
\hlinewd{1.1pt}
Variable & Value \\
\hline\\[-1.5mm]
\t{model} & \t{`QED'} \\[1.5mm]
\multirow{1}{5.2cm}{\centering\small{ \color{mygray} process 1:}}\\[2.5mm]
\t{InParticles} & \t{[`A']} \\[2.5mm]
\t{InParticles} & \t{[`f',`F']} \\[2.5mm]
\t{loops} & \t{1} \\[2.5mm]
\t{Parsel} & \t{[]} \\[2.5mm]
\t{LoopTec} & \t{3} \\[2.5mm]
\t{factor} & \t{`1'} \\[2.5mm]
\t{options} & \t{`onepi'} \\[1.5mm]
\multirow{1}{5.2cm}{\centering\small{ \color{mygray} process 2:}}\\[2.5mm]
\t{InParticles} & \t{[`f']} \\[2.5mm]
\t{InParticles} & \t{[`f']} \\[2.5mm]
\t{loops} & \t{1} \\[2.5mm]
\t{Parsel} & \t{[]} \\[2.5mm]
\t{LoopTec} & \t{3} \\[2.5mm]
\t{factor} & \t{`1'} \\[2.5mm]
\t{options} & \t{`onepi'} \\[2.5mm]
\multirow{1}{5.2cm}{\centering\small{ \color{mygray} Selection section:}}\\[2.5mm]
\t{FRinterLogic} & True \\[2.5mm]
\t{RenoLogic} & False \\[2.5mm]
\t{Draw} & False \\[2.5mm]
\t{Comp} & False \\[2.5mm]
\t{FinLogic} & False \\[2.5mm]
\t{DivLogic} & False \\[2.5mm]
\t{SumLogic} & False \\[2.5mm]
\t{MoCoLogic} & False \\[2.5mm]
\t{LoSpinors} & False \\[2.5mm]
\hlinewd{1.1pt}
\end{tabular}
\end{center}
\vspace{-5mm}
\end{normalsize}
\caption{Values of the variables of \t{Control.py} for the example in section \ref{sec:WI}.}
\label{tab:finalexample}
\end{table}
\normalsize
In order to prove eq. \ref{eq:provanda} with \ts{FeynMaster}, we need to consider the two processes depicted in eq. \ref{eq:pic}: the QED vertex and the fermion self-energy. Hence, we open and edit \t{Control.py} according to Table \ref{tab:finalexample} and we run \ts{FeynMaster}.\fn{{A \t{Control.py} file adapted to this example was included inside the folder containing the \ts{FeynRules} model for QED, under the name \t{Control-Example-5.3.py}.}}
Then, we move to the directory \t{1-AfF} inside \t{dirPro} (meanwhile generated), we copy the notebook that lies there to a different directory (say, \t{dirPro}), and we rename it \t{Notebook-Global.nb}. This is going to be the notebook where we shall combine the information of both processes. We open it, and delete most of the lines there: in a first phase, we only want to load the general files. So it must look like this:
\vs{1mm}
\begin{small}
\begin{addmargin}[8mm]{0mm}
\begin{Verbatim}[commandchars=\\\{\}]
\textcolor{mygray}{In[1]:=} << FeynCalc`
\textcolor{mygray}{In[2]:=} dirNuc = "(...)"; 
\textcolor{mygray}{In[3]:=} dirFey = "(...)"; 
\textcolor{mygray}{In[4]:=} Get["Feynman-Rules-Main.m", Path -> \{dirFey\}] 
\textcolor{mygray}{In[5]:=} Get["FunctionsOneLoop.m", Path -> \{dirNuc\}] 
\end{Verbatim}
\end{addmargin}
\end{small}
\vs{-1mm}
Now, we want to load the first process. To do so, we write:
\vs{1mm}
\begin{small}
\begin{addmargin}[8mm]{0mm}
\begin{Verbatim}[commandchars=\\\{\}]
\textcolor{mygray}{In[6]:=} dirHome = "(...)"; 
\textcolor{mygray}{In[7]:=} SetDirectory[dirHome];
\textcolor{mygray}{In[8]:=} << Helper.m;
\textcolor{mygray}{In[9]:=} Get["Definitions.m", Path -> \{dirNuc\}];
\textcolor{mygray}{In[10]:=} Get["Finals.m", Path -> \{dirNuc\}]; 
\end{Verbatim}
\end{addmargin}
\end{small}
\vs{-1mm}
where \t{dirHome} should be set to the directory corresponding to \t{1-AfF}. Next, we define new variables: \t{X0} as $\Gamma$ of eq. \ref{eq:provanda}, with the above-mentioned momentum definitions, and \t{X1} as the whole left-hand side of eq. \ref{eq:provanda}.
\vs{1mm}
\begin{small}
\begin{addmargin}[8mm]{0mm}
\begin{Verbatim}[commandchars=\\\{\}]
\textcolor{mygray}{In[11]:=} X0 = res1 /. \{p1 -> p2 - p3, q1 -> p2, q2 -> -p3\};
\textcolor{mygray}{In[12]:=} X1 = Contract[X0, FV[p2 - p3, -J1]];
\end{Verbatim}
\end{addmargin}
\end{small}
\vs{-1mm}
We now load the second process:
\vs{1mm}
\begin{small}
\begin{addmargin}[8mm]{0mm}
\begin{Verbatim}[commandchars=\\\{\}]
\textcolor{mygray}{In[13]:=} dirHome = "(...)"; 
\textcolor{mygray}{In[14]:=} SetDirectory[dirHome];
\textcolor{mygray}{In[15]:=} << Helper.m ;
\textcolor{mygray}{In[16]:=} Get["Definitions.m", Path -> \{dirNuc\}];
\textcolor{mygray}{In[17]:=} Get["Finals.m", Path -> \{dirNuc\}]; 
\end{Verbatim}
\end{addmargin}
\end{small}
\vs{-1mm}
where \t{dirHome} should now be set to the directory corresponding to \t{2-ff}. From this, and recalling that the default momentum of a self-energy is \t{p1} (cf. Table \ref{tab:FCconv}), we can obtain the right-hand side of eq. \ref{eq:provanda} by writing:
\vs{1mm}
\begin{small}
\begin{addmargin}[8mm]{0mm}
\begin{Verbatim}[commandchars=\\\{\}]
\textcolor{mygray}{In[18]:=} Y0a = res1 /. {p1 -> p2}; 
\textcolor{mygray}{In[19]:=} Y0b = res1 /. {p1 -> p3};
\textcolor{mygray}{In[20]:=} Y1 = (ee*(Y0a - Y0b) // DiracSimplify) // Simplify;
\end{Verbatim}
\end{addmargin}
\end{small}
\vs{-1mm}
Finally, we prove the Ward identity by showing that both sides of eq. \ref{eq:provanda} are equal:
\vs{1mm}
\begin{small}
\begin{addmargin}[8mm]{0mm}
\begin{Verbatim}[commandchars=\\\{\}]
\textcolor{mygray}{In[21]:=} WI = (Y1 - X1) // Simplify;
\textcolor{mygray}{In[22]:=} CheckWI = MyPaVeReduce[WI]
\end{Verbatim}
\end{addmargin}
\end{small}
\vs{-1mm}
which yields 0, thus completing the proof.

{
	\section{Brief comparison with other softwares}
	\label{sec:Comparison}
	
	\n We should not end this paper without providing an (even if brief) explanation of how \ts{FeynMaster} relates to other frameworks which also study to particle physics processes at the one-loop level. Among the most closely related to \ts{FeynMaster}, we select five: \ts{GRACE}~\cite{Belanger:2003sd}, \ts{FormCalc}~\cite{Hahn:1998yk}, \ts{GoSam}~\cite{Cullen:2011ac,Cullen:2014yla}, \ts{MadGraph5_aMC@NLO}~\cite{Alwall:2014hca} and \ts{NLOCT}~\cite{Degrande:2014vpa}.\fn{For a review comparing different softwares, see \cite{Harlander:1998dq}.}
	
	\n Before entering the comparison, let us consider a preliminary aspect. As we insisted throughout the paper, one of the major advantages of the present software is the flexible manipulation of intermediate and final results, due to the user-friendly features of \ts{FeynCalc} and \t{\ts{Mathematica}}. Now, if it is true that a software based on \t{\ts{Mathematica}} is flexible, it is also true that softwares based on other frameworks (such as \t{\ts{FORM}}~\cite{Kuipers:2012rf}) can be much faster. In fact, if one is interested in obtaining the numerical result for a one-loop process with hundreds of diagrams in a well known model, other softwares are certainly preferable to \ts{FeynMaster}. If, however, one wants to use a slightly different model, or ascertain some properties of the analytical results, or make sure to have no problems with interfaces between different softwares, or simply obtain a printed list of the Feynman rules, then \ts{FeynMaster} is an interesting alternative. In this sense, the purpose of \ts{FeynMaster} is not compete with the five aforementioned programs, but rather to enrich the flexibility of \ts{FeynCalc} by creating a unified platform around it that performs a vast list of tasks.
	
	%\n In order to help the comparison, recall from the Introduction that such list was:
	%\vs{1.0mm}
	%\begin{center}
	%\quad a) generation of Feynman rules; \qquad b) generation and drawing of Feynman diagrams;\\
	%c) generation of amplitudes; \hs{3mm} d) loop calculations; \hs{3mm} e) algebraic calculations; \hs{3mm} f) renormalization.
	%\end{center}
	%\vs{1.0mm}
	%%
	
	\n Let us now consider the referred programs. \ts{GRACE} is a very robust software that aims at numerically computing cross sections. It generates Feynman rules from a model, generates and draws Feynman diagrams, and writes the differential cross section in a \t{\ts{Fortran}} file. It is not restricted to 2 particles in the final state; it employs other softwares for the numerical integration and event generation; it also addresses renormalization, and calculates counterterms in the on-shell renormalization scheme. However, it cannot calculate one-loop processes that do not exist at tree-level, such as $h\to\gamma\gamma$. And although it performs algebraic calculations, it does not seem to allow a large margin for the user to manipulate analytical expressions.
	
	\n \ts{FormCalc} is a program that calculates one-loop Feynman diagrams based on both \t{\ts{FORM}} and \t{\ts{Mathematica}}. It combines the speed of the former and the vast amount of instructions of the latter. It gets the amplitudes from \ts{FeynArts}~\cite{Hahn:2000kx} and generates an output to numerically evaluate the one-loop integrals with \ts{LoopTools}. Although \ts{FormCalc} is excellent for fast results, it lacks the flexibility to perform simple manipulations involving Dirac algebra or Lorentz contractions, which are very useful to check properties of the processes (see, for example, section \ref{sec:FT}).
	
	\n \ts{GoSam} is designed to an automated calculation of one-loop amplitudes for multi-particle processes. Like \ts{FeynMaster}, it uses \ts{QGRAF} to generate the Feynman diagrams in symbolic form, and also uses both \t{\ts{Python}} and \t{feynmf} to draw the diagrams. Like \ts{GRACE}, \ts{GoSam} produces a \t{\ts{Fortran}} file required to perform the evaluation of one-loop matrix elements. But unlike \ts{FeynMaster}, it uses helicity amplitudes formalism, and makes renormalization for QCD corrections only. And although it generates one-loop amplitudes, it seems to be mostly focused on numerical evaluation, and not to allowing a manipulation of intermediate or final expressions.
	
	\n \ts{MadGraph5_aMC@NLO} is a very powerful framework for parton shower simulations and event analysis. It computes tree-level and one-loop amplitudes for arbitrary processes. However, it is not complete in one-loop processes of general models. For example, the one-loop corrections are restricted to QCD in the Standard Model. Hence, although it has some points in common with \ts{FeynMaster}, its focus is essentially different.
	
	\n \ts{NLOCT} determines the UV counterterms for any Lagrangean in an automated way. It requires an interaction with both \ts{FeynRules} and \ts{FeynArts}, and works only in the Feynman gauge. It not only obtains the Feynman rules for the counterterms, but also computes the analytical expressions for these counterterms in the on-shell renormalization scheme. Hence, concerning renormalization, \ts{NLOCT} is more powerful than \ts{FeynMaster}, which only calculates (automatically) analytical expressions in $\overline{\text{MS}}$ and does so only by considering a sequence of processes (see section \ref{sec:FullReno}). It should be clear, however, that it is certainly possible to use \ts{FeynMaster} to renormalize a full model like the Standard Model in the on-shell scheme, although such procedure will not be fully automatic. 
	
	\n In conclusion, although some of softwares considered here are more powerful than \ts{FeynMaster} in some specific tasks (e.g., \ts{FormCalc} for fast results, \ts{NLOCT} for an on-shell renormalization of different models), \ts{FeynMaster} exploits the strength of \ts{FeynCalc} to allow a practical manipulation of the analytical expressions. In addition --- it should not be forgotten ---, \ts{FeynMaster} condenses in itself a multiplicity of tasks, so that there is no need to convert (and hence there are no obstacles in converting) between different softwares. All this makes \ts{FeynMaster} a truly flexible framework. 
}

\section{Summary}
\label{sec:Summary}

\n We introduced the new software \ts{FeynMaster}, designed to perform several tasks of particle physics studies in a flexible and consistent way. We described in detail how to install it and how to use it, and gave some examples. {Finally, we compared it with some other relevant softwares, and concluded that its main advantages stem from its practical and multifaceted character.}

\n For a quick first usage of \ts{FeynMaster}, the user should follow this sequence of steps:
\vs{1.0mm}
\begin{center}
\begin{addmargin}[8mm]{0mm}
1) Make sure you have installed \t{\ts{Python}}, \t{\ts{Mathematica}} and \LaTeX \,, on the one hand, and \ts{FeynRules}, \ts{QGRAF} and \ts{FeynCalc}, on the other;\\[2mm]
2) Download \ts{FeynMaster} in \url{https://porthos.tecnico.ulisboa.pt/FeynMaster/};\\[2mm]
3) Extract the downloaded file and place the resulting folder in a directory at will;\\[2mm]
4) Edit the files \t{RUN-FeynMaster} and \t{Control.py} as explained in section \ref{sec:direct};\\[2mm]
5) Run \t{RUN-FeynMaster}.
\end{addmargin}
\end{center}
This should generate and open 4 PDF files relative to QED: the Feynman rules for the non-renormalized interactions, the Feynman rules for the counterterms
interactions, the Feynman diagram for the 1-loop vacuum polarization, and a document containing not only the expressions for the vacuum polarization, but also the expression for the associated counterterm in $\overline{\text{MS}}$.

%automatically  generates Feynman rules, generates and draws Feynman diagrams, generates amplitudes, performs both loop and algebraic calculations, and fully renormalizes models

%Explicar convenÃ§Ãµes de i's nas expressÃµes.

\section*{Acknowledgements}

\n Both authors are very grateful to António P. Lacerda, who kicked off the entire program. We also thank Vladyslav Shtabovenko, Paulo Nogueira and Augusto Barroso for useful discussions concerning \ts{FeynCalc}, \ts{QGRAF} and renormalization, respectively; Maximilian Löschner for bringing the \t{feynmf} package to our attention; Miguel P. Bento and Patrick Blackstone for testing the program; { Darius Jur\v{c}iukonis for} {a careful reading of the manuscript;} João P. Silva for the suggestion of the name `\ts{FeynMaster}', as well as for a careful reading of the manuscript. D.F. is also grateful to Isabel Fonseca for many useful suggestions concerning \t{\ts{Python}} and to Sofia Gomes for a suggestion regarding the printing of the Feynman rules. Both authors are supported by projects
CFTP-FCT Unit 777 (UID/FIS/00777/2013 and UID/FIS/00777/2019), and PTDC/FIS-PAR/ 29436/2017, which are partially funded through POCTI (FEDER), COMPETE, QREN and EU. D.F. is also supported by the Portuguese \textit{Funda\c{c}\~ao para a Ci\^encia e Tecnologia} under the project SFRH/BD/135698/2018.

% ====================================== Text ends here

\end{document}

%% file: PackagesnStyles.tex
\usepackage{geometry}	
\usepackage{setspace}

\everydisplay{}

\usepackage{tocloft}
\usepackage[T1]{fontenc}

\usepackage{indentfirst}

\usepackage{dashrule}

\usepackage{listings}
\usepackage[flushleft]{threeparttable}

\usepackage{fancyvrb}
\usepackage{xcolor}

\usepackage{mathtools}

\usepackage{titlesec}
\makeatletter
\def\hlinewd#1{%
	\noalign{\ifnum0=`}\fi\hrule \@height #1 %
	\futurelet\reserved@a\@xhline}
\makeatother

\usepackage{verbatim}
 
\usepackage{hyperref}
\hypersetup{colorlinks=false}
\usepackage{array}

\usepackage[latin9,utf8]{inputenc} % latin9 input encoding
\usepackage[T1]{fontenc}
\usepackage{ae}

\let\oldcdot\cdot 
\usepackage{breqn} 
\let\cdot\oldcdot 
\usepackage{etoolbox}
\patchcmd{\thebibliography}{\chapter*}{\section*}{}{}

\let\OLDthebibliography\thebibliography
\renewcommand\thebibliography[1]{
  \OLDthebibliography{#1}
  \setlength{\itemsep}{5pt}
  \setstretch{1.0}
}

\usepackage{multirow}

\usepackage{longtable}

\usepackage{scrextend}
\usepackage{ragged2e}
\usepackage{cancel}
\usepackage{float,hypcap}
\usepackage{changepage,titlesec}
\usepackage{sectsty}
\usepackage{soul}
\usepackage{bbold}
\usepackage{slashed}
\usepackage{tabularx}
\usepackage{amsmath,amsfonts,amssymb}
\usepackage{amsmath,bbm,latexsym,amssymb}
\usepackage{braket}
\usepackage{enumerate}
\usepackage{booktabs,hyperref}
\usepackage{bigints}
\usepackage{cite}

\usepackage{graphicx}
\usepackage{caption}
\usepackage{subcaption}
\usepackage{feynmp-auto}

\usepackage{color}
\definecolor{orange}{rgb}{1,0.5,0}
\definecolor{uglyblue}{RGB}{95,158,160}
\definecolor{mygray}{RGB}{129,129,129}

%% file: Definitions.tex
\def\be{\begin{equation}}
\def\ee{\end{equation}}
\def\ba{\begin{alignedat}}
\def\ea{\end{alignedat}}
\def\bea{\begin{eqnarray}}
\def\eea{\end{eqnarray}}
\newcommand{\bs}{\begin{subequations}}
\newcommand{\es}{\end{subequations}}
\def\bd{\begin{dmath}} 
\def\ed{\end{dmath}} 

\def\vs{\vspace}
\def\hs{\hspace}

\definecolor{mycolor}{RGB}{165,111,29}
\def\n{\noindent}
\def\fn{\footnote}
\def\t{\text}

\def\t{\texttt}
\def\ts{\textsc}
\def\FM{\textsc{FeynMaster}}

\def\LT{\textsc{LoopTools}}